\documentclass[12pt,reqno]{amsart}
\usepackage{amsfonts,color}
\usepackage{mathrsfs}
\usepackage{graphicx} 
\usepackage{epsfig}
\usepackage{amsmath, amsthm}
\usepackage{amssymb}
\usepackage{subfigure,psfrag}
\usepackage[capposition=top]{floatrow}
\usepackage[left=2.5cm, right=2.5cm, top=2.5cm, bottom=2.5cm]{geometry}

\numberwithin{equation}{section}
\date{}

\theoremstyle{plain}
\newtheorem{theorem}{Theorem}[section]
\newtheorem{proposition}[theorem]{Proposition}
\newtheorem{lemma}[theorem]{Lemma}
\newtheorem{corollary}[theorem]{Corollary}
\newtheorem{remark}{Remark}[section]
\theoremstyle{definition}
\newtheorem{definition}[theorem]{Definition}
\newtheorem{example}[theorem]{Example}
\numberwithin{equation}{section}

\begin{document}

\def\calL{\mathcal{L}}
\def\calG{\mathcal{G}}
\def\calD{\mathcal{D}}
\def\calJ{\mathcal{J}}
\def\calM{\mathcal{M}}
\def\calN{\mathcal{N}}
\def\calO{\mathcal{O}}
\def\calA{\mathcal{A}}
\def\calS{\mathcal{S}}
\def\calP{\mathcal{P}}
\def\calU{\mathcal{U}}
\def\calK{\mathcal{K}}
\def\frakgl{\mathfrak{gl}}
\def\frako{\mathfrak{o}}
\def\fraku{\mathfrak{u}}
\def\frakg{\mathfrak{g}}
\def\frakso{\mathfrak{so}}
\def\fraksl{\mathfrak{sl}}
\def\fraksp{\mathfrak{sp}}
\def\fraksu{\mathfrak{su}}
\def\F{\mathbb{F}}
\def\R{\mathbb{R}}
\def\N{\mathbb{N}}
\def\C{\mathbb{C}}
\def\M{\mathbb{M}}
\def\H{\mathbb{H}}
\def\P{\mathbb{P}}
\def\al{\alpha}
\def\be{\beta}
\def\p{\partial}
\def\n{\, | \, }
\def\ti{\tilde}
\def\a{\alpha}
\def\r{\rho}
\def\l{\lambda}
\def\hcalG {\hat{\mathcal{G}}}
\def\diag{{\rm diag \/ }}
\def\det{{\rm det \/ }}
\def\sp{{\rm span \/ }}
\def\rd{{\rm d\/}}
\def\K{\nabla}
\def\g{\gamma}
\def\Re{{\rm Re\/}}
\def\a{\alpha}
\def\b{\beta}
\def\d{\delta}
\def\D{\triangle}
\def\e{\epsilon}
\def\g{\gamma}
\def\G{\Gamma}
\def\K{\nabla}
\def\l{\lambda}
\def\L{\Lambda}
\def\n{\,\vert\,}
\def\o{\theta}
\def\w{\omega}
\def\W{\Omega}
\def\ca{{\mathcal{A}}}
\def\cd{{\mathcal{D}}}
\def\cf{{\mathcal{F}}}
\def\cg{{\mathcal{G}}}
\def\ch{{\mathcal{H}}}
\def\ck{{\mathcal{K}}}
\def\cl{{\mathcal{L}}}
\def\cL{{\mathcal{L}}}
\def\cm{{\mathcal{M}}}
\def\cn{{\mathcal{N}}}
\def\co{{\mathcal{O}}}
\def\cp{{\mathcal{P}}}
\def\cs{{\mathcal{S}}}
\def\ct{{\mathcal{T}}}
\def\cu{{\mathcal{U}}}
\def\cv{{\mathcal{V}}}
\def\cx{{\mathcal{X}}}
\def\li{\langle}
\def\ri{\rangle}
\def\n{\ \vert\ }
\def\tr{{\rm tr}}
\def\bs{\bigskip}
\def\ms{\medskip}
\def\ss{\smallskip}
\def\hb{\hfil\break\vskip -12pt}

\def\di{$\diamond$}
\def\ni{\noindent}
\def\ti{\tilde}
\def\p{\partial}
\def\Re{{\rm Re\/}}
\def\Im{{\rm Im\/}}
\def\I{{\rm I\/}}
\def\II{{\rm II\/}}
\def\diag{{\rm diag}}
\def\ad{{\rm ad}}
\def\Ad{{\rm Ad}}
\def\Iso{{\rm Iso}}
\def\Gr{{\rm Gr}}
\def\sgn{{\rm sgn}}

\def\rd{{\rm d\/}}

\def\R{\mathbb{R} }
\def\C{\mathbb{C}}
\def\H{\mathbb{H}}
\def\N{\mathbb{N}}
\def\Z{\mathbb{Z}}
\def\O{\mathbb{O}}
\def\F{\mathbb{F}}

\def\fg{\mathfrak{G}}

\newcommand{\beg}{\begin{example}}
\newcommand{\eeg}{\end{example}}
\newcommand{\bthm}{\begin{theorem}}
\newcommand{\ethm}{\end{theorem}}
\newcommand{\bprop}{\begin{proposition}}
\newcommand{\eprop}{\end{proposition}}
\newcommand{\bcor}{\begin{corollary}}
\newcommand{\ecor}{\end{corollary}}
\newcommand{\blem}{\begin{lemma}}
\newcommand{\elem}{\end{lemma}}
\newcommand{\bca}{\begin{cases}}
\newcommand{\eca}{\end{cases}}
\newcommand{\brem}{\begin{remark}}
\newcommand{\erem}{\end{remark}}
\newcommand{\bpm}{\begin{pmatrix}}
\newcommand{\epm}{\end{pmatrix}}
\newcommand{\bbm}{\begin{bmatrix}}
\newcommand{\ebm}{\end{bmatrix}}
\newcommand{\bvm}{\begin{vmatrix}}
\newcommand{\evm}{\end{vmatrix}}
\newcommand{\bdefn}{\begin{definition}}
\newcommand{\edefn}{\end{definition}}
\newcommand{\bsub}{\begin{subtitle}}
\newcommand{\esub}{\end{subtitle}}
\newcommand{\bex}{\begin{example}}
\newcommand{\eex}{\end{example}}
\newcommand{\ben}{\begin{enumerate}}
\newcommand{\een}{\end{enumerate}}
\newcommand{\bpf}{\begin{proof}}
\newcommand{\epf}{\end{proof}}

\newcommand{\balign}{\begin{align}}
\newcommand{\ealign}{\end{align}}
\newcommand{\baligns}{\begin{align*}}
\newcommand{\ealigns}{\end{align*}}
\newcommand{\beq}{\begin{equation}}
\newcommand{\eeq}{\end{equation}}
\newcommand{\beqs}{\begin{equation*}}
\newcommand{\eeqs}{\end{equation*}}
\newcommand{\beqa}{\begin{eqnarray}}
\newcommand{\eeqa}{\end{eqnarray}}
\newcommand{\beqas}{\begin{eqnarray*}}
\newcommand{\eeqas}{\end{eqnarray*}}

\def\pdo{$\psi$do}

\def\calA{{\mathcal A}}
\def\calB{{\mathcal B}}
\def\calD{{\mathcal D}}
\def\calF{{\mathcal F}}
\def\calG{{\mathcal G}}
\def\calJ{{\mathcal J}}
\def\calK{{\mathcal K}}
\def\calL{{\mathcal L}}
\def\calM{{\mathcal M}}
\def\calN{{\mathcal N}}
\def\calO{{\mathcal O}}
\def\calP{{\mathcal P}}
\def\calR{{\mathcal R}}
\def\calS{{\mathcal S}}
\def\calU{{\mathcal U}}
\def\calV{{\mathcal V}}

\def\li{\langle}
\def\ri{\rangle}

\def\frakP{{\mathfrak{P}}}

\def\half{\frac{1}{2}}
\def\Tr{{\rm Tr\/}}
\def\nkdv{$n\times n$ KdV}

\def \a {\alpha}
\def \b {\beta}
\def \d {\delta}
\def \D {\triangle}
\def \e {\epsilon}
\def \g {\gamma}
\def \G {\Gamma}
\def \K {\nabla}
\def \l {\lambda}
\def \L {\Lambda}
\def \n {\,\vert\,}
\def \N {\,\Vert\,}
\def \o {\theta}
\def\w{\omega}
\def\W{\Omega}
\def \s {\sigma}
\def \S {\Sigma}

\def\ca{{\mathcal {A}}}
\def\cC{{\mathcal {C}}}
\def\cg{{\mathcal {G}}}
\def\ci{{\mathcal {I}}}
\def\ck{{\mathcal {K}}}
\def\cl{{\mathcal {L}}}
\def\cm{{\mathcal {M}}}
\def\cn{{\mathcal {N}}}
\def\co{{\mathcal {O}}}
\def\cp{{\mathcal {P}}}
\def\cs{{\mathcal {S}}}
\def\ct{{\mathcal {T}}}
\def\cu{{\mathcal {U}}}
\def\ch{{\mathcal {H}}}

\def\R{{\mathbb{R}}}
\def\C{{\mathbb{C}}}
\def\H{{\mathbb{H}}}
\def\Z{{\mathbb{Z}}}

\def\Re{{\rm Re\/}}
\def\Im{{\rm Im\/}}
\def\tr{{\rm tr\/}}
\def\Id{{\rm Id\/}}
\def\I{{\rm I\/}}
\def\II{{\rm II\/}}
\def\li{\leftrangle}
\def\ri{rightrangle}
\def\id{{\rm Id}}
\def\gk{\frac{G}{K}}
\def\uk{\frac{U}{K}}

\def\p{\partial}
\def\li{\langle}
\def\ri{\rangle}
\def\ti{\tilde}
\def\i{\/ \rm i }
\def\j{\/ \rm j }
\def\k{\/ \rm k}
\def\n {\ \vert\ }
\def\bu{$\bullet$}
\def\ni{\noindent}
\def\ii{{\rm i\,}}

\def\bs{\bigskip}
\def\ms{\medskip}
\def\ss{\smallskip}

\title[]{Anatomy of a $q$-generalization of the Laguerre/Hermite Orthogonal Polynomials}
 
\author{Chuan-Tsung Chan$^\dagger$ \and Hsiao-Fan Liu$^{\ddagger}$}
\address{}
\dedicatory{$^\dagger$ Department of Applied Physics, Tunghai University\\
$^{\ddagger}$Institute of Mathematics, Academia Sinica\\
$^\dagger$ ctchan@go.thu.edu.tw, $^{\ddagger}$ liuhf@gate.sinica.edu.tw}

\date{\today} 

\begin{abstract}


We study a $q$-generalization of the classical Laguerre/Hermite orthogonal 
polynomials. Explicit results include: the recursive coefficients, matrix elements 
of generators for the Heisenberg algebra, and the Hankel determinants. The 
power of quadratic relation is illustrated by comparing two ways of calculating 
recursive coefficients. Finally, we derive a $q$-deformed version of the Toda 
equations for both $q$-Laguerre/Hermite ensembles, and check the compatibility 
with the quadratic relation.

\end{abstract}

\maketitle

\lineskip=0.25cm
\section{Introduction}\

One special property about the orthogonal polynomials \cite{IS11} is that the matrix representations 
of the Heisenberg algebra are usually very simple. For instance, the famous three-term
recurrence relation \cite{W07, BW10,C13} among orthogonal polynomials
\beq
x p_n=a_{n+1} p_{n+1}+b_n p_n+a_n p_{n-1}
\eeq 
implies that the matrix form of the position operator, $x$, with respect to the orthogonal polynomial 
basis, is tri-diagonal for all kinds of weight functions. 
In addition, for many classical weight functions, one can show that both translation 
($\frac{d}{dx}$) and dilation ($x \frac{d}{dx}$) operators are also tri-diagonal, such as 
the Laguerre and Hermite weight functions \cite{ISV87, SS94, 33C45, KSZ11, 39A13}. 
In this paper, we shall extend these classical 
examples to a more general setting, which we call generalized $q$-Laguerre/Hermite 
ensembles, and give explicit results for the representation of the Heisenberg algebra. 
By {\it generalized}, we mean that, in addition to the exponential functions, we also multiply 
the {\it standard} Laguerre/Hermite weights by a polynomial prefactor. Or equivalently, 
we add a logarithmic term to the polynomial potential (see the definitions in Sec. $2$).

Admittedly, the study of the Laguerre/Hermite ensembles might seem uninteresting for the 
apparent linearity features. However, from physical point of view, these two sets of orthogonal 
polynomials lie at the heart of fundamental physics. In particular, they are associated 
with the quantum systems of three dimensional hydrogen atom (associated Laguerre polynomials) 
and one dimensional simple harmonic oscillator (Hermite polynomials). Given the 
mysterious connection between these two systems, \cite{Fujikawa:1996sw}, it might be worthwhile to examine 
the physical relevance of the quadratic relation \cite{BW10} between these two ensembles (see Sec. $2$). 
Furthermore, $q$-deformation are often 
considered as certain microscopic/quantum modifications of the existing physical models. 
Hence we would like to extend the classical scenarios to a possible $q$-generalizations. 

To summarize, our new findings in this paper include:
\begin{itemize}
\item A systematic study of the generalized  $q$-Laguerre and  $q$-Hermite
               ensembles, including explicit solutions of the recursive coefficients, 
               their orthogonal polynomial representations of the
               the Heisenberg algebra, and their corresponding Hankel determinants 
               (at $\kappa=1$).
\item We elucidate the power of the quadratic relation in comparing two ways 
               of solving recursive coefficients for the $q$-Hermite ensemble.
\item We find  $q$-generalized Toda equations for the 
               recursive coefficients of the $q$-Laguerre/Hermite ensembles, and check 
               their compatibility with the quadratic relation.  
\end{itemize}

This paper is organized as follows: after briefly reviewing the basic ideas of the quadratic relations for the classical 
$q$-deformed Laguerre/Hermite ensembles in section $2$, we give explicit 
solutions of the dilation matrix elements and the recursive coefficients of the 
$q$-Laguerre/Hermite ensembles in section $3$. Section $4$ is devoted to the 
computations of the Hankel determinants for both ensembles. In section $5$, we 
propose possible $q$-deformed Toda equations for the $q$-Hermite and $q$-Laguerre 
ensembles, which are based on the $\kappa$-deformation of the $q$-derivative matrix 
elements and that of the recursive coefficients. A brief summary and future outlook 
conclude this paper at section $6$.

\section{The Quadratic Relations for the Laguerre/Hermite Orthogonal Polynomials}

\subsection{Review of the quadratic relation among recursive coefficients for the orthogonal polynomials associated with the classical 
             Laguerre/Hermite weights}\label{sec:1}\
\vspace{0.5cm}   



Given the classical Laguerre weight defined as 
\beq\label{}
v^{(\a)}(x;\kappa):= x^\a \exp(-\kappa x),\quad 0 \leq \kappa, \quad -1<\a, \quad 0 \leq x,
\eeq
we can compute the orthonormal polynomials $p_n^{(\a)}(x;\kappa)$ as
\beq\label{jc}
\int_0^\infty p_m^{(\a)}(x;\kappa)p_n^{(\a)}(x;\kappa) v^{(\a)}(x;\kappa) dx =\d_{mn} 
\eeq
through Gram-Schmidt process.

Similarly, from the classical Hermite weight, 
\beq\label{}
\w^{(\a)}(x;\kappa):=|x|^{2\a+1}\exp(-\kappa^2x^2), \quad 0 \leq \kappa, \quad x \in \R,
\eeq 
we obtain associated orthonormal polynomials $P_n^{(\a)}(x,\kappa)$ as
\beq\label{}
\int_{-\infty}^\infty P_m^{(\a)}(x;\kappa)P_n^{(\a)}(x;\kappa)\w^{(\a)}(x;\kappa) dx=\d_{mn}.
\eeq

The quadratic relation among these two sets of orthonormal polynomials
is based on a simple connection between the Lagurre and Hermite 
weights. Namely, 
\beq\label{ba}
\w^{(\a)}(x;\kappa)=|x|v^{(\a)}(x^2;\kappa^2).
\eeq 
One immediate consequence of Eq. \eqref{ba} is that, the orthonormal 
polynomials $P_n^{(\a)}(x;\kappa)$ can be expressed in 
terms of the orthogonal polynomials $p_n^{(\a)}(x;\kappa)$ as follows,
\beq\label{}
P_{2n}^{(\a)}(x;\kappa)=p_n^{(\a)}(x^2;\kappa^2), \quad
P_{2n+1}^{(\a)}(x;\kappa)=x p_n^{(\a+1)}(x^2;\kappa^2).
\eeq

Having established this fundamental relation, we can then derive many 
useful consequences. In this paper, we shall focus on quadratic relation for 
the recursive coefficients. 

The set of orthonormal polynomials associated with any weight function can
be viewed as a complete set of basis for the function space. Hence, it
induces a natural realization of the Heisenberg algebra, $[\frac{d}{dx},x]=1.$
In particular, the matrix elements of the position operator consist of the 
three-term recursive coefficients among orthonormal polynomials. In 
the case of the Laguerre weight, it is given as
\beq\label{}
x p_n^{(\a)}(x;\kappa)=a_{n+1}^{(\a)}(\kappa) p_{n+1}^{(\a)}(x;\kappa)+b_n^{(\a)}(\kappa) p_n^{(\a)}(x;\kappa)
+a_n^{(\a)}(\kappa) p_{n-1}^{(\a)}(x;\kappa),
\eeq 
and in the case of the Hermite weight, we have
\beq\label{bb}
x P_n^{(\a)}(x;\kappa)=A_{n+1}^{(\a)}(\kappa) P_{n+1}^{(\a)}(x;\kappa)+A_n^{(\a)}(\kappa) P_{n-1}^{(\a)}(x;\kappa).
\eeq 
By computing $x^2P_n^{(\a)}(x,\kappa)$ in two ways, we obtain the quadratic 
relation among the two sets of recursive coefficients:
\begin{align}
a_n^{(\a)}(\kappa^2)&=A_{2n}^{(\a)}(\kappa)A_{2n-1}^{(\a)}(\kappa), \label{2-1-1}\\
b_n^{(\a)}(\kappa^2)&=\left(A_{2n+1}^{(\a)}(\kappa)\right)^2+\left(A_{2n}^{(\a)}(\kappa)\right)^2.\label{2-1-2}
\end{align}

\subsection{On the quadratic relation between generalized $q$-Laguerre/Hermite ensembles}\
\vspace{0.5cm}

In this paper, we take generalized $q$-Hermite and $q$-Laguerre ensembles
as an illustrative example of the quadratic relation. We consider the generalized $q$-Laguerre weight ($0 \leq \kappa < \frac{1}{q}$) 
\beq 
v^{(\a)}(x;\kappa,q) := |x|^\a (q\kappa x; q)_\infty=|x|^\a \prod_{l=0}^\infty (1-q^{l+1}\kappa x),
\eeq and the generalized $q$-Hermite 
weight 
\beq
\w^{(\a)}(x;\kappa,q) = |x|^{2\a+1} (q^2\kappa^2x^2;q^2)_\infty=|x| v^{(\a)}(x^2;\kappa^2,q^2).
\eeq

Given the orthonormal polynomials of the $q$-Laguerre ensemble $p_n^{(\a)}(x;\kappa,q)$, 
we can express the orthonormal polynomials of the $q$-Hermite ensemble $P_n^{(\a)}(x;\kappa,q)$ 
as follows:
\beqa\label{ga}
P_{2n}^{(\a)}(x;\kappa,q)&=&\sqrt{\frac{1+q}{2}}p_n^{(\a)}(x^2;\kappa^2,q^2)  \mbox{ (even, } \deg=2n),\\
P_{2n+1}^{(\a)}(x;\kappa,q)&=&\sqrt{\frac{1+q}{2}} x p_n^{(\a+1)}(x^2;\kappa^2,q^2) \mbox{ (odd, } \deg=2n+1).
\eeqa

One can easily check that $P_{2n}^{(\a)}(x;\kappa,q)$ and $P_{2n+1}^{(\a)}(x;\kappa,q)$ satisfy the orthonormal 
conditions, for instance,
\begin{align}
&\int_{-1}^1 P_{2m}^{(\a)}(x;\kappa,q) P_{2n}^{(\a)}(x;\kappa,q)\w^{(\a)}(x;\kappa,q) d_q x\nonumber\\
=&2(1-q) \sum_{k=0}^\infty P_{2m}^{(\a)}(q^k;\kappa,q) P_{2n}^{(\a)}(q^k;\kappa,q)\w^{(\a)}(q^k;\kappa,q) q^k\nonumber\\
=&2(1-q)\left(\frac{1+q}{2}\right)\sum_{k=0}^\infty p_m^{(\a)}(q^{2k};\kappa^2,q^2) p_n^{(\a)}(q^{2k};\kappa^2,q^2) q^k v^{(\a)}(q^{2k};\kappa^2,q^2) q^k\nonumber\\
=&\int_0^1 p_m^{(\a)}(x;\kappa^2,q^2) p_n^{(\a)}(x;\kappa^2,q^2) v^{(\a)}(x;\kappa^2,q^2) d_q x=\d_{mn}.
\end{align}
\begin{align}
&\int_{-1}^1 P_{2m+1}^{(\a)}(x;\kappa,q) P_{2n+1}^{(\a)}(x;\kappa,q)\w^{(\a)}(x;\kappa,q) d_q x\nonumber\\
=&2(1-q) \sum_{k=0}^\infty P_{2m+1}^{(\a)}(q^k;\kappa,q) P_{2n+1}^{(\a)}(q^k;\kappa,q)\w^{(\a)}(q^k;\kappa,q) q^k\nonumber\\
=&2(1-q)\left(\frac{1+q}{2}\right)\sum_{k=0}^\infty q^k p_m^{(\a+1)}(q^{2k};\kappa^2,q^2) q^k p_n^{(\a+1)}(q^{2k};\kappa^2,q^2) q^k v^{(\a)}(q^{2k};\kappa^2,q^2) q^k\nonumber\\
=&(1-q^2)\sum_{k=0}^\infty p_m^{(\a+1)}(q^{2k};\kappa^2,q^2) p_n^{(\a+1)}(q^{2k};\kappa^2,q^2) v^{(\a+1)}(q^{2k};\kappa^2,q^2) q^{2k}\nonumber\\
=&\int_0^1 p_m^{(\a+1)}(x;\kappa^2,q^2) p_n^{(\a+1)}(x;\kappa^2,q^2) v^{(\a+1)}(x;\kappa^2,q^2) d_q x=\d_{mn}.
\end{align}
The $P_{2m}^{(\a)}$-$P_{2m+1}^{(\a)}$ orthogonality is trivial due to the even parity of 
the generalized $q$-Hermite weight.

Similar to the classical cases Eqs. \eqref{2-1-1}, \eqref{2-1-2}, there exists a correspondence between 
recursive coefficients associated with the generalized $q$-Laguerre and $q$-Hermite ensembles. 
\bthm\label{thm:2-1}
The recursive coefficients associated with $q$-generalized Laguerre and Hermite ensembles satisfying 
the following relations:
\beqa
a_n^{(\a)}(\kappa^2,q^2)&=&A_{2n}^{(\a)}(\kappa,q)A_{2n-1}^{(\a)}(\kappa,q), \label{bc} \\
b_n^{(\a)}(\kappa^2,q^2)&=&\left(A_{2n+1}^{(\a)}(\kappa,q)\right)^2+\left(A_{2n}^{(\a)}(\kappa,q)\right)^2, \label{bd}\\
a_n^{(\a+1)}(\kappa^2,q^2)&=&A_{2n+1}^{(\a)}(\kappa,q)A_{2n}^{(\a)}(\kappa,q),\label{be}\\
b_n^{(\a+1)}(\kappa^2,q^2)&=&\left[A_{2n+2}^{(\a)}(\kappa,q)\right]^2+\left[A_{2n+1}^{(\a)}(\kappa,q)\right]^2.\label{bf}
\eeqa
\ethm
\bpf
\begin{align*}
&x^2 P_{2n}^{(\a)}(x;\kappa,q)\\
&=x\left[A_{2n+1}^{(\a)}(\kappa,q)P_{2n+1}^{(\a)}(x;\kappa,q)+A_{2n}^{(\a)}(\kappa,q)P_{2n-1}^{(\a)}(x;\kappa,q)\right]\\
                                   &=A_{2n+1}^{(\a)}(\kappa,q)\left[A_{2n+2}^{(\a)}(\kappa,q)P_{2n+2}^{(\a)}(x;\kappa,q)+A_{2n+1}^{(\a)}(\kappa,q)P_{2n-1}^{(\a)}(x;\kappa,q)\right]\\
                                   &+A_{2n}^{(\a)}(\kappa,q)\left[A_{2n}^{(\a)}(\kappa,q)P_{2n}^{(\a)}(x;\kappa,q)+A_{2n-1}^{(\a)}(\kappa,q)P_{2n-2}^{(\a)}(x;\kappa,q)\right]\\
                                   &=\left[A_{2n+1}^{(\a)}(\kappa,q)A_{2n+2}^{(\a)}(\kappa,q)\right]P_{2n+2}^{(\a)}(x;\kappa,q)+\left[\left(A_{2n+1}^{(\a)}(\kappa,q)\right)^2+\left(A_{2n}^{(\a)}(\kappa,q)\right)^2\right]P_{2n}^{(\a)}(x;\kappa,q)\\
                                   &+[A_{2n-1}^{(\a)}(\kappa,q)A_{2n}^{(\a)}(\kappa,q)]P_{2n-2}^{(\a)}(x;\kappa,q).
\end{align*}
On the other hand, using the expression of Eq. \eqref{ga}, we have
\begin{align*}
&x^2 P_{2n}^{(\a)}(x;\kappa,q)\\
&=x^2 \sqrt{\frac{1+q}{2}}p_n^{(\a)}(x^2;\kappa^2,q^2)\\
&=\sqrt{\frac{1+q}{2}}\left[a_{n+1}^{(\a)}(\kappa^2,q^2)p_{n+1}^{(\a)}(x^2;\kappa^2,q^2)+b_n^{(\a)}(\kappa^2,q^2)p_n^{(\a)}(x^2;\kappa^2,q^2)+a_n^{(\a)}(\kappa^2,q^2)p_{n-1}^{(\a)}(x^2;\kappa^2,q^2)\right]\\
&=a_{n+1}^{(\a)}(\kappa^2,q^2)P_{2n+2}^{(\a)}(x;\kappa,q)+b_n^{(\a)}(\kappa^2,q^2)P_{2n}^{(\a)}(x;\kappa,q)+a_n^{(\a)}(\kappa^2,q^2)P_{2n-2}^{(\a)}(x;\kappa,q).
\end{align*}
By comparing the coefficients on both expressions, we get Eqs.\eqref{bc}, \eqref{bd}.

If we examine similar calculations for the odd $q$-Hermite orthonormal 
polynomials, we get
\begin{align*}
x^2 P_{2n+1}^{(\a)}&=[A_{2n+3}^{(\a)}A_{2n+2}^{(\a)}]P_{2n+3}^{(\a)}\\
&+[\left(A_{2n+2}^{(\a)}\right)^2+\left(A_{2n+1}^{(\a)}\right)^2]P_{2n+1}^{(\a)}+[A_{2n+1}^{(\a)}A_{2n}^{(\a)}]P_{2n-1}^{(\a)}.
\end{align*}
Alternatively,
\begin{align*}
&x^2 P_{2n+1}^{(\a)}(x;\kappa,q)\\
&=x^2 \sqrt{\frac{1+q}{2}}x p_n^{(\a+1)}(x^2;\kappa^2,q^2)\\
&=\sqrt{\frac{1+q}{2}}\left[a_{n+1}^{(\a+1)}(\kappa^2,q^2)p_{n+1}^{(\a+1)}(x^2;\kappa^2,q^2)+b_n^{(\a+1)}(\kappa^2,q^2)p_n^{(\a+1)}(x^2;\kappa^2,q^2)+a_n^{(\a+1)}(\kappa^2,q^2)p_{n-1}^{(\a+1)}(x^2;\kappa^2,q^2)\right]\\
&=a_{n+1}^{(\a+1)}(\kappa^2,q^2)P_{2n+3}^{(\a)}(x;\kappa,q)+b_n^{(\a+1)}(\kappa^2,q^2)P_{2n+1}^{(\a)}(x;\kappa,q)+a_n^{(\a+1)}(\kappa^2,q^2)P_{2n-1}^{(\a)}(x;\kappa,q).
\end{align*}
Thus, we have shown Eqs.\eqref{be}, \eqref{bf}.
\epf
Eliminating the recursive coefficients of the $q$-Laguerre orthonormal polynomials, 
$a_n^{(\a)},b_n^{(\a)}$, in both sets of the equation, we obtain
\beq\label{}
A_{2n}^{(\a)}(\kappa,q)A_{2n-2}^{(\a)}(\kappa,q)=A_{2n+1}^{(\a-1)}(\kappa,q)A_{2n}^{(\a-1)}(\kappa,q),
\eeq
and
\beq\label{}
\left(A_{2n+1}^{(\a)}(\kappa,q)\right)^2+\left(A_{2n}^{(\a)}(\kappa,q)\right)^2=\left(A_{2n+2}^{(\a-1)}(\kappa,q)\right)^2+\left(A_{2n+1}^{(\a-1)}(\kappa,q)\right)^2.
\eeq
We shall see how these quadratic relation will help us derive the explicit form of the recursive 
coefficients $A_n^{(\a)}(1,q)$ from the (easier) solutions of $a_n^{(\a)}(1,q^2), b_n^{(\a)}(1,q^2)$ in 
Sec. $3$. For the general $\kappa$ case, we check the compatibility between the quadratic relation 
and the evolution equations (w.r.t $\kappa$) in Sec. $5$.

\section{Explicit solutions of the Fourier and recursive coefficients for the $q$-Laguerre/Hermite ensembles}

In this section, we derive the matrix representations of the $q$-dilation 
operator in the generalized $q$-Laguerre/Hermite ensembles, and use 
it to solve for recursive coefficients of the orthonormal polynomials. Since 
we shall focus on the special case $\kappa=1$, we shall suppress the 
$\kappa$ (and $q$) dependences to make the equations simpler.

\subsection{Solution of the recursive coefficients for the generalized $q$-Laguerre ensemble (at $\kappa =1$)}\label{sec:3-1}\
\vspace{0.5cm}


\bthm\label{thm:3-1}
The matrix elements (Fourier coefficients) of the $q$-dilation operator with respect 
to the orthonormal $q$-Laguerre polynomials are given by
\beq
x \calD^x_q p_n^{(\a)}(x)=\left(\frac{1-q^n}{1-q}\right) p_n^{(\a)}(x)+\left[q^{\frac{-\a-1}{2}}(1-q)^{-1}(1-q^n)^{\frac{1}{2}}(1-q^{n+\a})^{\frac{1}{2}}\right] p_{n-1}^{(\a)}(x).
\eeq
\ethm

\bthm\label{thm:3-2}
The recursive coefficients for the orthonormal $q$-Laguerre polynomial associated 
with weight $v^{(\a)}(x;1,q)$ is given by
\beqa
\left(a_n^{(\a)}\right)^2&=&q^{2n+\a-1}(1-q^n)(1-q^{n+\a}) \geq 0,\\
b_n^{(\a)}&=&-q^{2n+\a}(1+q)+q^n(1+q^\a).
\eeqa
\ethm

\bpf[Proof of Theorems \ref{thm:3-1} and \ref{thm:3-2}]

To compute the recursive coefficients, $a_n^{(\a)}(q),b_n^{(\a)}(q)$, we first study the Fourier coefficients of the action of dilation
on the orthonormal $q$-Laguerre polynomials: 
\beq\label{} 
x\calD_q^x p_n^{(\a)}(x)=\sum_{j=0}^n p_j^{(\a)}(x) c_{jn}^{(\a)},
\eeq
where projection formula gives
\beqa\label{fa}
c_{jn}^{(\a)}&=&\int_{-1}^1 p_j^{(\a)}(x)[x \calD_q^x p_n^{(\a)}(x)]v^{(\a)}(x) d_qx\nonumber\\
                   &=&\frac{-1}{1-q} \int_{-1}^1p_j^{(\a)}(x) p_n^{(\a)}(qx) v^{(\a)}(x)d_q x\nonumber\\
                   &+&\frac{1}{1-q} \int_{-1}^1 p_j^{(\a)}(x) p_n^{(\a)}(x) v^{(\a)}(x)d_q x.
\eeqa
Using the Pearson relation for the generalized $q$-Laguerre weight,
 \beq\label{}
 v^{(\a)}\left(\frac{x}{q}\right)=q^{-\a}(1-x)v^{(\a)}(x),
 \eeq
we get
\beqa\label{3-1-1}
c_{jn}^{(\a)}&=&-\dfrac{q^{-(\a+1)}}{1-q} \int_{-1}^1p_j^{(\a)}\left(\dfrac{x}{q}\right) p_n^{(\a)}(x) v^{(\a)}(x)d_q x\nonumber\\
                   &+&\dfrac{q^{-(\a+1)}}{1-q} \int_{-1}^1 p_j^{(\a)}\left(\dfrac{x}{q}\right) p_n^{(\a)}(x) x v^{(\a)}(x)d_q x\nonumber\\
                   &+&\dfrac{1}{1-q} \d_{jn}.
\eeqa
From this expression, it is clear that $c_{jn}^{(\a)}=0$ if $j \leq n-2$. Thus, we have  
\beq\label{3-1-2} 
x\calD_q^x p_n^{(\a)}(x)=p_n^{(\a)}(x) c_{nn}^{(\a)}+p_{n-1}^{(\a)}(x) c_{n-1,n}^{(\a)}.
\eeq

From the coefficients of the orthonormal $q$-Laguerre polynomials:
\beq\label{} 
p_n^{(\a)}(x)=\g_n^{(\a)} \left(x^n + \eta_n^{(\a)} x^{n-1}+ \cdots\right), 
\eeq
we deduce the decomposition of the rescaled orthonormal polynomials,
\beqa
p_n^{(\a)}\left(\dfrac{x}{q}\right)&=&q^{-n}p_n^{(\a)}(x)+\dfrac{\eta_n^{(\a)}}{a_n^{(\a)}}q^{-n}(q-1)p_{n-1}^{(\a)}(x)+\cdots,\\
p_n^{(\a)}(qx)&=&q^np_n^{(\a)}(x)+\dfrac{\eta_n^{(\a)}}{a_n^{(\a)}}q^{n-1}(1-q)p_{n-1}^{(\a)}(x)+\cdots.
\eeqa

Substituting these results back to Eq.\eqref{fa}, we obtain
\beqa
c_{nn}^{(\a)}&=&\frac{1-q^n}{1-q} \mbox{  (independent of $\a$)}, \label{fb}\\
c_{n-1,n}^{(\a)}&=&-\frac{\eta_n^{(\a)}}{a_{n}^{(\a)}}q^{n-1}.\label{fd}
\eeqa

On the other hand, we can relate the Fourier coefficients, $c_{nn}^{(\a)}$ and $c_{n-1,n}^{(\a)}$ 
to the recursive coefficient $a_n^{(\a)}$ using the second integral expression \eqref{3-1-1} together 
with
\beq\label{}
x p_n^{(\a)}\left(\frac{x}{q}\right)=q^{-n}a_{n+1}^{(\a)}p_{n+1}^{(\a)}(x)+(\eta_n^{(\a)}q^{-n-1}-\eta_{n+1}^{(\a)}q^{-n})p_n^{(\a)}(x)+\cdots.
\eeq
The results are:
\beqa
c_{nn}^{(\a)}&=&-\frac{q^{-(\a+n+1)}}{1-q}+\frac{q^{-(\a+1)}}{1-q}(\eta_n^{(\a)}q^{-(n-1)}-\eta_{n+1}^{(\a)}q^{-n})+\frac{1}{1-q},\\ \label{fc}
c_{n-1,n}^{(\a)}&=&\frac{q^{-(\a+n)}}{1-q}a_n^{(\a)}.\label{fe}
\eeqa
By comparing two expressions for $c_{nn}^{(\a)}$, Eqs.\eqref{fb} and \eqref{fc}, we solve
\beq\label{}
\eta_n^{(\a)}=\frac{(q^{n+\a}-1)(1-q^n)}{1-q}.
\eeq
From the expressions for $c_{n-1,n}^{(\a)}$, Eqs.\eqref{fd} and \eqref{fe}, we solve
\beq\label{ha}
\left(a_n^{(\a)}\right)^2=q^{2n+\a-1}(1-q^n)(1-q^{n+\a}) \geq 0.
\eeq
Finally, the middle recursive coefficient $b_n^{(\a)}$ can be solved as 
\beq\label{hb}
b_n^{(\a)}=\eta_n^{(\a)}-\eta_{n+1}^{(\a)}=-q^{2n+\a}(1+q)+q^{n+\a}+q^n.
\eeq
\epf

\subsection{Solution of the recursive coefficients for the generalized $q$-Hermite ensemble (at $\kappa=1$)}\label{sec:3-2}\
\vspace{0.5cm}

In this section, we calculate the recursive coefficients for the orthonormal 
polynomials associated with generalized $q$-Hermite ensembles, $A_n^{(\a)}$. 
In order to derive a master equation for the recursive coefficients $A_n^{(\a)}$,
we first study the Fourier coefficients of the action of dilation on the orthonormal
$q$-Hermite polynomials. In this section, we set $\kappa=1$ and suppress the $q$ dependence. 

\bthm
The matrix elements (Fourier coefficients) of the $q$-dilation operator with respect to the orthonormal 
$q$-Hermite polynomials are given in terms of the recursive coefficients $A_n^{(\a)}$ as
\beqa
x\calD_q^x P_n^{(\a)}(x)&=&\sum_{j=0}^n P_j^{(\a)}(x) C_{jn}^{(\a)}\nonumber\\
&=&\frac{1-q^n}{1-q}P_n^{(\a)}(x)+\left[\frac{q^{n-2}(1+q)}{A_{n-1}^{(\a)}A_n^{(\a)}}\sum_{l=0}^{n-1}\left(A_l^{(\a)}\right)^2\right] P_{n-2}^{(\a)}(x).
\eeqa
\ethm

\bpf
The Fourier coefficients of the $q$-dilation operator can be computed by using projection formula,
\beq\label{}
C_{jn}^{(\a)}=\int_{-1}^1 P_j^{(\a)}(x)[x \calD_q^x P_n^{(\a)}(x)]\w^{(\a)}(x) d_qx
\eeq

Substituting the definition of the $q$-derivative $\calD_q^x P_n^{(\a)}$ (see \eqref{app:1}) and recalling the 
definition of the $q$-integration, we get
\beqa\label{ea}
C_{jn}^{(\a)}&=&-2 \sum_{l=0}^\infty P_j^{(\a)}(q^l)P_n^{(\a)}(q^{l+1}) \w^{(\a)}(q^l) q^l\nonumber\\
                   &+&2 \sum_{l=0}^\infty P_j^{(\a)}(q^l) P_n^{(\a)}(q^l)\w^{(\a)}(q^l) q^l\nonumber\\
                   &=&\dfrac{-1}{1-q} \int_{-1}^1P_j^{(\a)}(x) P_n^{(\a)}(qx) \w^{(\a)}(x)d_q x\nonumber\\
                   &+&\dfrac{1}{1-q} \int_{-1}^1 P_j^{(\a)}(x) P_n^{(\a)}(x) \w^{(\a)}(x)d_q x.
\eeqa

Upon shifting the index $\kappa$ and using the Pearson relation for the generalized $q$-Hermite
 weight
 \beq\label{}
 \w^{(\a)}\left(\frac{x}{q}\right)=q^{-(2\a+1)}(1-x^2)\w^{(\a)}(x),
 \eeq
we get
\beqa\label{eb}
C_{jn}^{(\a)}&=&-\dfrac{q^{-(2\a+2)}}{1-q} \int_{-1}^1P_j^{(\a)}\left(\frac{x}{q}\right) P_n^{(\a)}(x) \w^{(\a)}(x)d_q x\nonumber\\
                   &+&\dfrac{q^{-(2\a+2)}}{1-q} \int_{-1}^1 P_j^{(\a)}\left(\frac{x}{q}\right) P_n^{(\a)}(x) x^2 \w^{(\a)}(x)d_q x\nonumber\\
                   &+&\dfrac{1}{1-q} \d_{jn}.
\eeqa
From this expression, it is clear that $C_{jn}^{(\a)}=0$ if $j< n-2$. Thus, we have  
\beq\label{3-2-3} 
x\calD_q^x P_n^{(\a)}(x)=P_n^{(\a)}(x) C_{nn}^{(\a)}+P_{n-2}^{(\a)}(x) C_{n-2,n}^{(\a)}.
\eeq

To facilitate the computations, we need to invoke the structure of the orthonormal 
coefficients. Let
\beqs
P_n^{(\a)}(x)=\g_n^{(\a)} \left(x^n + \zeta_n^{(\a)} x^{n-2}+ \cdots\right),
\eeqs
we can decompose the rescaled orthonormal polynomials $P_n^{(\a)}\left(\dfrac{x}{q}\right)$,
$P_n^{(\a)}(qx)$ as follows: 
\beqa\label{}
P_n^{(\a)}\left(\dfrac{x}{q}\right)&=&\g_n^{(\a)} \left(q^{-n}x^n+\zeta_n^{(\a)}q^{-n+2}x^{n-2}+\cdots\right) \nonumber\\
                                  &=&q^{-n}\g_n^{(\a)}\left[(x^n+\zeta_n^{(\a)}x^{n-2}+\cdots) -\zeta_n^{(\a)}x^{n-2}
                                         +\zeta_n^{(\a)} q^{2} x^{n-2}+\cdots\right] \nonumber\\
                                  &=&q^{-n}P_n^{(\a)}(x)-\dfrac{q^{-n}(1-q^2)\zeta_n^{(\a)}}{A_{n-1}^{(\a)}A_n^{(\a)}}P_{n-2}^{(\a)}+\cdots,
\eeqa
\beqa\label{3-2-1}
P_n^{(\a)}(qx)&=&\g_n^{(\a)} \left(q^{n}x^n+\zeta_n^{(\a)}q^{n-2}x^{n-2}+\cdots\right)\nonumber\\
                                  &=&q^n\g_n^{(\a)}\left[(x^n+\zeta_n^{(\a)}x^{n-2}+\cdots) -\zeta_n^{(\a)}x^{n-2}
                                         +\zeta_n^{(\a)} q^{-2} x^{n-2}+\cdots\right]\nonumber\\
                                  &=&q^nP_n^{(\a)}(x)+\dfrac{q^{n-2}(1-q^2)\zeta_n^{(\a)}}{A_{n-1}^{(\a)}A_n^{(\a)}}P_{n-2}^{(\a)}+\cdots.
\eeqa
Substituting Eq. \eqref{3-2-1} back to Eq. \eqref{ea}, we obtain
\beqa
C_{nn}^{(\a)}&=&\frac{1-q^n}{1-q} \mbox{  ( independent of $\a$)},\label{3-2-2}\\
C_{n-2,n}^{(\a)}&=&\frac{-(1+q)q^{n-2}}{A_{n-1}^{(\a)}A_n^{(\a)}}\zeta_n^{(\a)}.
\eeqa
Note that 
\beq\label{}
\zeta_n^{(\a)}=-\sum_{l=0}^{n-1}\left(A_l^{(\a)}\right)^2,
\eeq
and
\beq\label{}
C_{n+1,n+1}^{(\a)}-C_{nn}^{(\a)}=q^n.
\eeq
\epf
It is worth emphasizing that, both Eqs. \eqref{fb}, \eqref{3-2-2} are just special 
cases of a universal feature of the $q$-dilation operator. We give a general proof 
without using Pearson relation in Appendix B.

Having derived the Fourier coefficients $C_{nn}^{(\a)}$ and $C_{n-2,n}^{(\a)}$ for 
$x\calD_q^x[P_n^{(\a)}(x)]$, we can use Eq.\eqref{eb} to derive a master equation for the 
recursive coefficients $A_n^{(\a)}$ defined in Eq.\eqref{bb}  with $\kappa=1$,
\beq\label{}
xP_n^{(\a)}(x)=A_{n+1}^{(\a)}P_{n+1}^{(\a)}(x)+A_n^{(\a)}P_{n-1}^{(\a)}(x).
\eeq
Namely,
\beq\label{}
xP_n^{(\a)}\left(\frac{x}{q}\right)=P_{n+1}^{(\a)}(x)[q^{-n}A_{n+1}^{(\a)}]
+P_{n-1}^{(\a)}(x)\left[q^{-n}A_n^{(\a)}-\frac{q^{-n}\zeta_n^{(\a)}(1-q^2)}{A_n^{(\a)}}\right]+\cdots,
\eeq
Eq.\eqref{eb} implies
\beq\label{}
\begin{array}{rcl}
1-q^n&=&(1-q)C_{nn}^{(\a)}\\
         &=&(1-q^{-n-2\a-2})
+q^{-n-2\a-2}\left[\left(A_{n+1}^{(\a)}\right)^2+\left(A_n^{(\a)}\right)^2-\zeta_n^{(\a)}(1-q^2)\right].
\end{array}
\eeq
By taking a first difference on both sides of the equation, $(1-q)(C_{n+1,n+1}^{(\a)}-C_{nn}^{(\a)})$, 
and define a shifted difference 
\beq\label{ed}
B_k^{(\a)}:=\left(A_k^{(\a)}\right)^2-q\left(A_{k-1}^{(\a)}\right)^2 \Leftrightarrow 
\left(A_n^{(\a)}\right)^2=\sum_{k=1}^n q^{n-k} B_k^{(\a)}, 
\eeq
with $A_0^{(\a)}=0, B_1^{(\a)}=\left(A_1^{(\a)}\right)^2$, we get
\beq\label{}
(1-q)q^{2n+2\a+3}=(q-1)+B_{n+2}^{(\a)}+B_{n+1}^{(\a)}+(1-q^2)\sum_{k=1}^n B_k^{(\a)}.
\eeq
By taking another difference $(1-q)(q^{2n+2\a+3}-q^{2n+2\a+1})$, and defining another shifted 
difference $C_n^{(\a)}:=B_{n+1}^{(\a)}-qB_n^{(\a)}.$ That is,
\beq\label{ec}
B_n^{(\a)}=q^{n-1}B_1^{(\a)}+\sum_{k=1}^{n-1} q^{n-1-k}C_k,
\eeq
we get
\beq\label{}
C_{n+1}^{(\a)}=-qC_n^{(\a)}-(1-q)^2(1+q)q^{2n+2\a+1}.
\eeq

The explicit form of $C_n^{(\a)}$ can be solved by iteration, and is given by
\beq\label{}
C_n^{(\a)}=(-q)^{n-1}C_1^{(\a)}-(1-q)^2(-1)^n q^{n+2\a+1}[1-(-q)^{n-1}],
\eeq
where $C_1^{(\a)}=B_2^{(\a)}-q B_1^{(\a)}=\left(A_2^{(\a)}\right)^2-2q\left(A_1^{(\a)}\right)^2$.

Substituting the explicit form of $C_n^{(\a)}$, and performing the summation, Eq.\eqref{ec}, 
we get for $m \geq 1$,
\beqa\label{}
B_{2m}^{(\a)}&=&-q^{2m-1}\left(A_1^{(\a)}\right)^2+q^{2m-2}\left(A_2^{(\a)}\right)^2-q^{2m+2\a+1}(1-q)(1-q^{2m-2}),\\
B_{2m-1}^{(\a)}&=&q^{2m-2}\left(A_1^{(\a)}\right)^2-q^{2m+2\a-1}(1-q)(1-q^{2m-2}).
\eeqa
Substituting the explicit forms of $B_n^{(\a)}$, and performing the summation, \eqref{ed}, we
get
\beqa
\left(A_{2m}^{(\a)}\right)^2&=&mq^{2m-2}\left(A_2^{(\a)}\right)^2+q^{2m+2\a}[(1-q^{2m})-m(1-q^2)], \label{ee}\\
\left(A_{2m-1}^{(\a)}\right)^2&=&q^{2m-2}\left(A_1^{(\a)}\right)^2+(m-1)q^{2m-3}\left(A_2^{(\a)}\right)^2\nonumber\\
&+&\frac{q^{2m+2\a}}{1+q}[(1-q^{2m-2})-(m-1)(1-q)]. \label{ef}
\eeqa

\subsection{The power of the quadratic relation}\
\vspace{0.5cm}

To show the power of the quadratic relation, we first use the solutions of 
the recursive coefficients of the $q$-Laguerre orthonormal polynomials 
$a_n^{(\a)},b_n^{(\a)}$, Eqs.\eqref{ha}, \eqref{hb} to deduce the recursive coefficients of the 
$q$-Hermite polynomials. 
\bthm
The recursive coefficients for the orthonormal polynomials of the $q$-Hermite ensemble are
\beqa
\left[A_{2n}^{(\a)}(1,q)\right]^2 &=&q^{2n+2\a}(1-q^{2n}),\\
\left[A_{2n+1}^{(\a)}(1,q)\right]^2&=&q^{2n}(1-q^{2n+2\a+2}).
\eeqa
\ethm
\bpf
First of all, we have learned that, using Eqs. \eqref{bc}, \eqref{bd},
\begin{align*}
\left(A_{2n}^{(\a)}(q)\right)^2\left(A_{2n+1}^{(\a)}(q)\right)^2&=\left(a_n^{(\a+1)}(q^2)\right)^2\\
&=q^{4n+2\a}(1-q^{2n})(1-q^{2n+2\a+2}),
\end{align*}
and
\begin{align*}
\left(A_{2n}^{(\a)}(q)\right)^2+\left(A_{2n+1}^{(\a)}(q)\right)^2&=b_n^{\a}(q^2)\\
&=-q^{4n+2\a}(1+q^{2n})+q^{2n}(q^{2\a}+1).
\end{align*}

This implies that $\left(A_{2n}^{(\a)}\right)^2$and $\left(A_{2n+1}^{(\a)}\right)^2$ can 
be solved from the quadratic equation,
\beq\label{}
z^2-b_n^{(\a)}(q^2)z+\left(a_n^{(\a+1)}(q^2)\right)^2=0,
\eeq
and two roots are
\beq\label{}
z_{\pm}^{(\a)}(q)=\frac{b_n^{(\a)}(q^2)\pm \sqrt{\left(b_n^{(\a)}(q^2)\right)^2-4\left(a_n^{(\a+1)}(q^2)\right)^2}}{2}.
\eeq
Substituting the solutions for $a_n^{(\a+1)}(q^2)$ and $b_n^{(\a)}(q^2)$, Eqs.\eqref{ha},\eqref{hb}, we get
\begin{align*}
z_{\pm}^{(\a)}(q)&=\frac{-q^{4n+2\a}(1+q^2)+q^{2n}(1+q^{2\a}) \pm [q^{4n+2\a}(1-q^2)+q^{2n}(1-q^{2\a})]}{2}\\
&=\bca
      -q^{4n+2\a+2}+q^{2n} \ \,\quad(+)\\
      -q^{4n+2\a}+q^{2n+2\a} \quad (-)
     \eca
\end{align*}
By taking $\a \rightarrow -\frac{1}{2}$, 
\beq\label{}
z_{\pm}^{(-\frac{1}{2})}(q) \rightarrow 
     \bca
      -q^{4n+1}+q^{2n} \rightarrow A_{2n+1}^{(-\frac{1}{2})} \\
      -q^{4n-1}+q^{2n-1} \rightarrow A_{2n}^{(-\frac{1}{2})} .
     \eca
\eeq
We deduce that 
\beqa
\left(A_{2n}^{(\a)}\right)^2&=&q^{2n+2\a}(1-q^{2n}) \quad \Rightarrow \left(A_2^{(\a)}\right)^2=q^{2\a+2}(1-q^2)\label{eg}\\
\left(A_{2n+1}^{(\a)}\right)^2&=&q^{2n}(1-q^{2n+2\a+2}) \quad \Rightarrow \left(A_1^{(\a)}\right)^2=1-q^{2\a+2}. \label{eh}
\eeqa
\epf
From this second approach, it is clear that not only we can easily derive the recursive coefficients of the generalized 
$q$-Hermite orthonormal polynomials, but also we get the {\it initial values} $A_1^{(\a)},A_2^{(\a)}$ 
for free. One can now check by substituting the {\it initial} values $A_1^{(\a)},A_2^{(\a)}$ into Eqs.\eqref{ee},
\eqref{ef}, which are solutions from (laborious) successive reduction, that we obtain consistent results as 
derived from the quadratic relation.

\section{Hankel determinants and its $q$-generalization}

\subsection{Hankel determinant for the $q$-Laguerre ensemble (for $\kappa=1$)}\

\vspace{0.5cm}
In this section, we shall compute the Hankel determinant associated with
the generalized $q$-Laguerre ensemble, which is defined as the determinant 
of the moments associated with the weight function.
\beq\label{}
\mu_k^{(\a)}:=\int_0^1 x^k v^{(\a)}(x,q)d_q x,
\eeq
\beq\label{}
\D_n^{(\a)}:=\left|\begin{array}{cccc}
                     \mu_0^{(\a)} & \mu_1^{(\a)} & \cdots & \mu_{n-1}^{(\a)} \\
                     \mu_1^{(\a)} & \mu_2^{(\a)} & \cdots & \mu_n^{(\a)} \\
                     \vdots & \vdots &  & \vdots \\
                     \mu_{n-1}^{(\a)} & \mu_n^{(\a)} & \cdots & \mu_{2n-2}^{(\a)}
                     \end{array}\right|, \quad \D_0^{(\a)}:=1.
\eeq
\beq\label{}
\ti\D_n^{(\a)}:=\left|\begin{array}{ccccc}
                     \mu_0^{(\a)} & \mu_1^{(\a)} & \cdots & \mu_{n-2}^{(\a)} &\mu_n^{(\a)} \\
                     \mu_1^{(\a)} & \mu_2^{(\a)} & \cdots & \mu_{n-1}^{(\a)} &\mu_{n+1}^{(\a)} \\
                     \vdots & \vdots &  & \vdots &\vdots \\
                     \mu_{n-1}^{(\a)} & \mu_n^{(\a)} & \cdots & \mu_{2n-3}^{(\a)} &\mu_{2n-1}^{(\a)}
                     \end{array}\right|, \quad\ti\D_0^{(\a)}:=1.
\eeq

In addition, the recursive coefficients for the orthonormal polynomials can 
be expressed in terms of the Hankel determinants as
\beq\label{}
\left(a_n^{(\a)}\right)^2=\frac{\D_{n+1}^{(\a)}\D_{n-1}^{(\a)}}{\left(\D_n^{(\a)}\right)^2},\quad
b_n^{(\a)}=\frac{\ti\D_{n+1}^{(\a)}}{\D_{n+1}^{(\a)}}-\frac{\ti\D_n^{(\a)}}{\D_n^{(\a)}}.
\eeq
Defining the ratio among the Hankel determinants of adjacent rank,
\beq\label{}
r_n^{(\a)}:=\frac{\D_n^{(\a)}}{\D_{n-1}^{(\a)}}, \quad
r_1^{(\a)}=\frac{\D_1^{(\a)}}{\D_0^{(\a)}}=\mu_0^{(\a)}=\frac{1}{(\g_0^{(\a)})^2},
\eeq    
we see that 
\beq\label{}
\left(a_n^{(\a)}\right)^2=\frac{r_{n+1}^{(\a)}}{r_n^{(\a)}}, \quad \mbox{and }
r_n^{(\a)}=r_1^{(\a)} \prod_{k=1}^{n-1} \left(a_k^{(\a)}\right)^2.
\eeq
Now we can substitute the solutions of the recursive coefficients, Eq.\eqref{ha}, 
for $\left(a_k^{(\a)}\right)^2$ and we get
\beqa
r_n^{(\a)}&=&r_1^{(\a)} q^{\sum_{j=1}^{n-1}(2j+\a-1)}\left(\prod_{l=1}^{n-1} (1-q^l)(1-q^{l+\a})\right)\nonumber\\
                &=&r_1^{(\a)} q^{(n-1)(n+\a-1)}\left(\prod_{l=1}^{n-1}(1-q^l)(1-q^{\a+l})\right).
\eeqa
Next, we compute the Hankel determinant
\beqa
\D_n^{(\a)}&=&\D_1^{(\a)} \prod_{k=2}^n r_n^{(\a)}\nonumber\\
                 &=&\D_1^{(\a)}\left(r_1^{(\a)}\right)^{n-1}q^{\sum_{j=2}^n(j-1)(j+\a-1)}\prod_{m=1}^{n-1} \prod_{l=1}^m \left((1-q^l)(1-q^{l+\a})\right)\nonumber\\
                 &=&\left(\D_1^{(\a)}\right)^n q^{\frac{n(n-1)(2n+3\a-1)}{6}}\prod_{j=1}^{n-1} (1-q^j)^{n-j}(1-q^{j+\a})^{n-j}.
\eeqa
The value of zeroth moment, $\mu_0^{(0)}=\D_1^{(0)}$, for the $q$-Laguerre weight is given in Appendix C.

\subsection{Hankel determinant for the $q$-Hermite ensemble (for $\kappa=1$)}\

\vspace{0.5cm}
Following the same definitions, we compute the Hankel determinant associated 
with the $q$-Hermite ensemble. We recall the solutions of the recursive coefficients,
Eqs.\eqref{eg},\eqref{eh},
\begin{align*}
\left(A_{2l}^{(\a)}(q)\right)^2&=q^{2(l+\a)}(1-q^{2l}),\\
\left(A_{2l-1}^{(\a)}(q)\right)^2&=q^{2(l-1)}(1-q^{2l+2\a}),
\end{align*}
and define the ratios among adjacent Hankel determinants, $R_n^{(\a)}:=\dfrac{\D_n^{(\a)}}{\D_{n-1}^{(\a)}}$,
 we get
\beqa\label{ja}
 R_{2m}^{(\a)}&=&R_1^{(\a)}\prod_{l=1}^{2m-1}\left(A_l^{(\a)}\right)^2\nonumber\\
                      &=&R_1^{(\a)}\left\{\prod_{j=1}^{m-1} \left(A_{2j}^{(\a)}\right)^2\right\}\left\{\prod_{l=1}^m \left(A_{2l-1}^{(\a)}\right)^2\right\}\nonumber\\
                      &=&R_1^{(\a)} q^{2\sum_{j=1}^{m-1}(j+\a)}\left(\prod_{k=1}^{m-1}(1-q^{2k}) \right)q^{2\sum_{l=1}^m (l-1)} \left(\prod_{l=1}^m (1-q^{2l+2\a})\right)\nonumber\\
                      &=&R_1^{(\a)}q^{2(m+\a)(m-1)}\prod_{k=1}^{m-1} (1-q^{2k}) \prod_{l=1}^m(1-q^{2l+2\a}).
\eeqa
Similarly, 
\beqa\label{jb}
 R_{2m-1}^{(\a)}&=&R_1^{(\a)}\prod_{l=1}^{2m-2}\left(A_l^{(\a)}\right)^2\nonumber\\
                      &=&R_1^{(\a)}\left\{\prod_{j=1}^{m-1} \left(A_{2j}^{(\a)}\right)^2\right\}\left\{\prod_{l=1}^{m-1} \left(A_{2l-1}^{(\a)}\right)^2\right\}\nonumber\\
                      &=&R_1^{(\a)} q^{(m+2\a)(m-1)}\left(\prod_{k=1}^{m-1}(1-q^{2k}) \right)q^{(m-2)(m-1)} \left(\prod_{l=1}^{m-1} (1-q^{2l+2\a})\right)\nonumber\\
                      &=&R_1^{(\a)}q^{2(m-1)(m+\a-1)}\prod_{k=1}^{m-1} (1-q^{2k})(1-q^{2k+2\a}).
\eeqa
 From the results of Eqs.\eqref{ja}, \eqref{jb}, we can then derive explicit expressions for the Hankel
 determinants as a finite product. For the even case,
\beqa
 \D_{2j}^{(\a)}&=&\D_1^{(\a)}\prod_{k=2}^{2j} R_k^{(\a)}\nonumber\\
                     &=&\D_1^{(\a)}\left(\prod_{k=1}^j R_{2k}^{(\a)}\right)\left(\prod_{l=1}^{j-1} R_{2l+1}^{(\a)}\right)\nonumber\\
                     &=&\left(\D_1^{(\a)}\right)^{2j} q^{\frac{j(j-1)(4j+6\a+1)}{3}}\left(\prod_{k=1}^{j-1} (1-q^{2k})^{2j-2k}\right)
                     \left(\prod_{l=1}^j (1-q^{2l+2\a})^{2j-2l+1}\right).
\eeqa
 The odd case can be easily computed
 \begin{align*}
 \D_{2j-1}^{(\a)}&= \frac{\D_{2j}^{(\a)}}{R_{2j}^{(\a)}}\\
                        &=\left(\D_1^{(\a)}\right)^{2j-1} q^{\frac{j-1}{3}[4j^2+(6\a-5)j-6\a]}\left( \prod_{k=1}^{j-1} (1-q^{2k})^{2j-2k-1}(1-q^{2k+2\a})^{2j-2k} \right).
 \end{align*}
As a consistent check, we specialize our results to $\a=0$ and $\a=-\frac{1}{2}$.\\

$\a=0:$\\
\beqa
R_{2j}^{(0)}&=&R_1^{(0)} q^{2j(j-1)}(1-q^{2j})\prod_{k=1}^{j-1} (1-q^{2k})^2,\\
R_{2j-1}^{(0)}&=&R_1^{(0)} q^{2(j-1)^2} \prod_{k=1}^{j-1} (1-q^{2k})^2,\\
\D_{2j}^{(0)}&=&\left(\D_1^{(0)}\right)^{2j} q^{\frac{j(j-1)(4j+1)}{3}} \prod_{k=1}^{j-1}\left((1-q^{2k})^{4j-4k+1}\right)^{(1-q^{2j})},\\
\D_{2j-1}^{(0)}&=&\left(\D_1^{(0)}\right)^{2j-1} q^{\frac{j(j-1)}{3}(4j-5)} \prod_{k=1}^{j-1}(1-q^{2k})^{4j-4k-1}.
\eeqa
 
$\a=-\frac{1}{2}:$\\
\beqa
R_{2j}^{(-\frac{1}{2})}&=&R_1^{(-\frac{1}{2})} q^{(2j-1)(j-1)}\prod_{k=1}^{2j-1} (1-q^k),\\
R_{2j-1}^{(-\frac{1}{2})}&=&R_1^{(-\frac{1}{2})} q^{(2j-3)(j-1)} \prod_{k=1}^{2j-2} (1-q^k),\\
\D_{2j}^{(-\frac{1}{2})}&=&\left(R_1^{(-\frac{1}{2})}\right)^{2j} q^{\frac{2j(j-1)(2j-1)}{3}} \prod_{k=1}^{2j-1}\left((1-q^k)^{2j-k}\right),\\
\D_{2j-1}^{(-\frac{1}{2})}&=&\left(R_1^{(-\frac{1}{2})}\right)^{2j-1} q^{\frac{(j-1)(2j-1)(2j-3)}{3}} \prod_{k=1}^{2j-2}(1-q^k)^{2j-k-1}.
\eeqa
The value of zeroth moment, $\mu_0^{(-\frac{1}{2})}=\D_1^{(-\frac{1}{2})}$, for the $q$-Hermite weight is given in Appendix C.

\section{$q$-Generalization of the Toda equations from $\kappa$-deformation of the $q$-Laguerre/Hermite ensembles}

\subsection{$q$-Difference equations for the recursive coefficients of the $q$-Laguerre
polynomials}\
\vspace{0.5cm}


In this section, we study the $q$-difference equations describing the $\kappa$ dependence 
of the recursive coefficients $a_n^{(\a)}(\kappa),b_n^{(\a)}(\kappa)$ associated with the 
generalized $q$-Laguerre ensemble. In the classical case, such equations correspond to 
the Lax pair formulation of the Toda equations \cite{F74}. Hence, our results provide a $q$-generalization 
of the classical Toda equation. To achieve this, we introduce the Fourier expansion (w.r.t $\kappa$ variable) of the $q$-derivative 
on the $q$-Laguerre orthonormal polynomials,
\beq\label{ia}
\calD_q^\kappa p_n^{(\a)}(x,\kappa)=\sum_{j=0}^n p_j^{(\a)}(x,\kappa) \xi_{j n}^{(\a)}(\kappa).
\eeq
Recalling the definition of the $q$-derivative, \eqref{app:1}, we can also transform this expansion 
formula as a $q$-shifting relation ($\l:=(1-q)\kappa$):
\beq\label{}
p_n^{(\a)}(x,q\kappa)= p_n^{(\a)}(x,\kappa)[1-\l \xi_{nn}^{(\a)}(\kappa)]-\l \sum_{j=0}^{n-1} p_j^{(\a)}(x,\kappa)\xi_{j n}^{(\a)}(\kappa).
\eeq

Next, we compute the $q$-derivative with respect to the $\kappa$ variable on the 
action of position operator, $x p_n^{(\a)}(x,\kappa)$, in two ways: 

We first compute the $q$-derivative with respect to $\kappa$ on the recursive relation,
\beqa\label{ib}
\calD_q^\kappa[x p_n^{(\a)}(x,\kappa)]
&=&\calD_q^\kappa[a_{n+1}^{(\a)}(\kappa)p_{n+1}^{(\a)}(x,\kappa)+b_n^{(\a)}(\kappa)p_n^{(\a)}(x,\kappa)+a_n^{(\a)}(\kappa)p_{n-1}^{(\a)}(x,\kappa)]\nonumber\\
&=&[\calD_q^\kappa a_{n+1}^{(\a)}+\xi_{n+1,n+1}^{(\a)}\bar a_{n+1}^{(\a)}]p_{n+1}^{(\a)}+[\calD_q^\kappa b_n^{(\a)}+\xi_{n,n+1}^{(\a)} \bar a_{n+1}^{(\a)}+\xi_{nn}^{(\a)} \bar b_n^{(\a)}]p_n^{(\a)}\nonumber\\
&+&[\calD_q^\kappa a_n^{(\a)}+\xi_{n-1,n-1}^{(\a)} \bar a_n^{(\a)}+\xi_{n-1,n}^{(\a)}\bar b_n^{(\a)}]p_{n-1}^{(\a)}+[\xi_{n-2,n-1}^{(\a)} \bar a_n^{(\a)}]p_{n-2}^{(\a)}.
\eeqa
Here $\xi_{mn}^{(\a)}$ are the Fourier coefficients (matrix elements) of Eq. \eqref{ia}, 
\beqs
\bar a_n^{(\a)}:=a_n^{(\a)}(q\kappa),\quad \bar b_n^{(\a)}:=b_n^{(\a)}(q\kappa)
\eeqs
are the rescaled recursive coefficients, and we suppress the dependence on $\kappa$ for 
simplicity. 

On the other hand, since $\calD_q^\kappa$ commutes with the position operator $x$, we first 
compute the $q$-derivative (w.r.t $\kappa$ variable) of the orthonormal polynomials and 
then apply the position operator.
\beqa\label{ic}
& &x \calD_q^\kappa[p_n^{(\a)}(x,\kappa)]\nonumber\\
&=&[\xi_{nn}^{(\a)} a_{n+1}^{(\a)}]p_{n+1}^{(\a)}+[\xi_{n-1,n}^{(\a)}  a_{n}^{(\a)}+\xi_{nn}^{(\a)} b_n^{(\a)}]p_n^{(\a)}\nonumber\\
&+&[\xi_{nn}^{(\a)} a_n^{(\a)}+\xi_{n-1,n}^{(\a)} b_{n-1}^{(\a)}]p_{n-1}^{(\a)}+[\xi_{n-1,n}^{(\a)} a_{n-1}^{(\a)}]p_{n-2}^{(\a)}.
\eeqa

By comparing the corresponding coefficients of each orthonormal polynomials as 
calculated in Eqs.\eqref{ib}, \eqref{ic}, we get the following set of relations:
\beqa
\calD_q^\kappa a_n^{(\a)}&=&\xi_{n-1,n-1}^{(\a)} a_n^{(\a)}-\xi_{nn}^{(\a)}\bar a_n^{(\a)},\label{id}\\
\calD_q^\kappa b_n^{(\a)}&=&\xi_{nn}^{(\a)}(b_n^{(\a)}-\bar b_n^{(\a)})+[\xi_{n-1,n}^{(\a)} a_n^{(\a)}-\xi_{n,n+1}^{(\a)} \bar a_{n+1}^{(\a)}],\\
\calD_q^\kappa a_n^{(\a)}&=&\xi_{n-1,n}^{(\a)}(b_{n-1}^{(\a)}-\bar b_n^{(\a)})+[\xi_{nn}^{(\a)} a_n^{(\a)}-\xi_{n-1,n-1}^{(\a)} \bar a_n^{(\a)}],\label{ie}\\
\xi_{n-2,n-1}^{(\a)}\bar a_n^{(\a)}&=&\xi_{n-1,n}^{(\a)}a_{n-1}^{(\a)}.
\eeqa
Note that the last result allows us to replace the rescaled recursive coefficients $\bar a_n^{(\a)}$
in terms of the Fourier coefficients and the unscaled recursive coefficients
\beq\label{5-1-2}
a_n^{(\a)}(q\kappa)=\frac{\xi_{n-1,n}^{(\a)}(\kappa)}{\xi_{n-2,n-1}^{(\a)}(\kappa)}a_{n-1}^{(\a)}(\kappa) 
=\frac{1-\l \xi_{n-1,n-1}^{(\a)}}{1-\l \xi_{nn}^{(\a)}}a_n^{(\a)}(\kappa),
\eeq
where the second equality of the above relation follows from Eq.\eqref{id}. We have also checked that Eqs.\eqref{id}, \eqref{ie} are compatible. 

Finally, after some manipulations, we 
get the coupled $q$-difference equations.
\beqa
\calD_q^\kappa a_n^{(\a)}(\kappa)&=&\frac{\xi_{n-1,n-1}^{(\a)}(\kappa)-\xi_{nn}^{(\a)}(\kappa)}{1-\l \xi_{nn}^{(\a)}(\kappa)} a_n^{(\a)}(\kappa) \label{toda:1}\\
\calD_q^\kappa b_n^{(\a)}(\kappa)&=&\frac{q}{1-q}\left[\left(\frac{a_n^{(\a)}(\kappa)}{1-\l \xi_{nn}^{(\a)}}\right)^2-\left(\frac{a_{n+1}^{(\a)}(\kappa)}{1-\l \xi_{n+1,n+1}^{(\a)}}\right)^2\right].\label{toda:2}
\eeqa

Our next task is to find an expression relating $\xi_{nn}^{(\a)}$ in terms of the recursive 
coefficients $a_n^{(\a)},b_n^{(\a)}$. By taking $q$-derivative w.r.t $\kappa$ variable on the orthonormal condition, and 
recalling the $q$-Leibniz rule, \eqref{app:2}, we can derive a master equation among these Fourier 
coefficients.
\beq\label{}
\calD_q^\kappa \left[\int_0^1 p_m^{(\a)}(x,\kappa)p_n^{(\a)}(x,\kappa) v^{(\a)}(x,\kappa)d_q x\right]=0.
\eeq
This implies for $m\leq n$,
\beq\label{}
(1-\l \xi_{mm}^{(\a)})\xi_{mn}^{(\a)}-\l \sum_{j=0}^{m-1} \xi_{jm}^{(\a)} \xi_{jn}^{(\a)}+\d_{mn} \xi_{nn}^{(\a)}
=\d_{m,n-1}\left[\frac{q}{1-q}\bar a_n^{(\a)}\right]+\d_{mn}\left[\frac{q}{1-q}\bar b_n^{(\a)}\right].
\eeq
From these results, we can exact useful information by specifying the value of $m$:
\begin{itemize}
\item[Case $1$:] $m< n-1 \Leftrightarrow m+2\leq n$\\
We find, by induction, 
$\xi_{mn}^{(\a)}(\kappa)=0$, if $m \leq n-2$. Consequently, there are only two terms in the 
$q$-derivative (w.r.t $\kappa$ variable) of the $q$-Laguerre orthonormal polynomials,
\beq
\calD_q^\kappa p_n^{(\a)}(x,\kappa)=p_n^{(\a)}(x,\kappa)\xi_{nn}^{(\a)}(\kappa)+p_{n-1}^{(\a)}\xi_{n-1,n}^{(\a)}(\kappa).
\eeq
\item[Case $2$:] $m=n-1$\\
In this case, we relate the two Fourier coefficients as follows:
\beq\label{5-1-1}
\xi_{n-1,n}^{(\a)}=\frac{\dfrac{q}{1-q}\bar a_n^{(\a)}}{1-\l \xi_{n-1,n-1}^{(\a)}}=\left(\frac{q}{1-q}\right)\frac{a_n^{(\a)}}{1-\l \xi_{nn}^{(\a)}}.
\eeq
\item[Case $3$:] $m=n$\\
By suitable rearrangements, we derive a recursive equation relating the diagonal Fourier coefficients
$\xi_{nn}^{(\a)}$ to the recursive coefficients $a_n^{(\a)},b_n^{(\a)}$ as follows:
\beq\label{if}
(1-\l \xi_{nn}^{(\a)})^2+\frac{(q \kappa \bar a_n^{(\a)})^2}{(1-\l \xi_{n-1,n-1}^{(\a)})^2}=1-q \kappa \bar b_n^{(\a)}.
\eeq
\end{itemize}

By using Eq.\eqref{5-1-2}, we can rewrite Eq.\eqref{if} as a quadratic equation for $(1-\l \xi_{nn}^{(\a)})^2$,
\beq\label{5-1-3}
(1-\l \xi_{nn}^{(\a)})^2+\frac{(q \kappa a_n^{(\a)})^2}{(1-\l \xi_{nn}^{(\a)})^2}=1-q \kappa \bar b_n^{(\a)}.
\eeq
From the solution of this equation, we then obtain an expression of $\xi_{nn}^{(\a)}$ in terms of $a_n^{(\a)}$ and 
$\bar b_n^{(\a)}$,
\beq\label{5-1-4}
\xi_{nn}^{(\a)}=\frac{1}{2(1-q)\kappa}\left\{2-\sqrt{(1-q\kappa \bar b_n^{(\a)})+\sqrt{1-2q\kappa \bar b_n^{(\a)}+4(q\kappa)^2\left[\left(\bar b_n^{(\a)}\right)^2-\left(a_n^{(\a)}\right)^2\right]}}\right\}.
\eeq
Substituting this expression back to Eqs.\eqref{toda:1}, \eqref{toda:2}, we then obtain a set of 
closed $q$-difference equations for the recursive coefficients, $a_n^{(\a)}$ and $b_n^{(\a)}$, of the $q$-generalized Laguerre ensemble.

\subsection{$q$-Difference equations for the recursive coefficients of the $q$-Hermite orthonormal polynomials}\
 \vspace{0.5cm}

In this section, we shall derive the $q$-difference equation for the recursive 
coefficients of the $q$-Hermite orthonormal polynomials. In order to achieve 
this, we need to introduce the Fourier coefficients of the $q$-derivative
of the $q$-Hermite orthonormal polynomials with respect to parameter $\kappa$,
\beq\label{db}
\calD_q^\kappa P_n^{(\a)}(x;\kappa) = \sum_{j=0}^n P_j^{(\a)}(x;\kappa) \Xi_{jn}^{(\a)}(\kappa), \quad (n-j \mbox{ is even}).
\eeq
Note that, by recalling the definition of the $q$-derivative, \eqref{app:1}, we 
can also transform the equation above into the Fourier expansion of the 
$q$-evolved $q$-Hermite orthonormal polynomials with respect to parameter $\kappa$,
\beq\label{dc}
P_n^{(\a)}(x;q\kappa) =[1-\l \Xi_{nn}^{(\a)}]P_n^{(\a)}(x;\kappa)-\l \sum_{j=0}^{n-1}P_j^{(\a)}(x;\kappa)
\Xi_{jn}^{(\a)}(\kappa).
\eeq 
Next, we compute 
the $\calD_q^\kappa$ derivative on the product $x P_n^{(\a)}(x;\kappa)$ in two ways.
\beqa\label{}
\calD_q^\kappa[x P_n^{(\a)}(x;\kappa)]&=&\calD_q^\kappa[A_{n+1}^{(\a)}(\kappa)P_{n+1}^{(\a)}(x;\kappa)+A_n^{(\a)}(\kappa)P_{n-1}^{(\a)}(x;\kappa)]\nonumber\\
&=&x [\sum_{j=0}^n P_j^{(\a)}(x;\kappa)\Xi_{jn}^{(\a)}(\kappa)]\nonumber\\
&=&\sum_{j=0}^n[A_{j+1}^{(\a)}(\kappa)P_{j+1}^{(\a)}(x;\kappa)+A_j^{(\a)}(\kappa)P_{j-1}^{(\a)}(x;\kappa)]\Xi_{jn}^{(\a)}(\kappa)].
\eeqa
By comparing the coefficients of $P_{n+1}^{(\a)}(x;\kappa)$ of the first and the third lines 
of the previous equation, we get the following results:
\beq\label{}
\calD_q^\kappa A_{n+1}^{(\a)}(\kappa)=  \Xi_{nn}^{(\a)}(\kappa) A_{n+1}^{(\a)}(\kappa)-\Xi_{n+1,n+1}^{(\a)}(\kappa) A_{n+1}^{(\a)}(q\kappa),
\eeq
which implies
\beq\label{toda:3}
\calD_q^\kappa A_n^{(\a)}(\kappa) =\frac{\Xi_{n-1,n-1}^{(\a)}(\kappa)-\Xi_{nn}^{(\a)}(\kappa)}{1+\kappa(q-1)\Xi_{nn}^{(\a)}(\kappa)}A_n^{(\a)}(\kappa),
\eeq
and
\beq\label{5-2-3}
\frac{A_n^{(\a)}(q\kappa)}{1-\l \Xi_{n-1,n-1}^{(\a)}(\kappa)} =\frac{A_n^{(\a)}(\kappa)}{1-\l\Xi_{n,n}^{(\a)}(\kappa)}.
\eeq
Our next task is to derive a set of algebraic equations for the Fourier coefficients 
$\Xi_{jn}^{(\a)}(\kappa)$. By taking the $q$-derivative (w.r.t $\kappa$) on the orthonormal 
condition
\beq\label{}
\calD_q^\kappa \left(\int_{-1}^1 P_m^{(\a)}(x;\kappa) P_n^{(\a)}(x;\kappa) \w^{(\a)}(x;\kappa) d_q x\right)=0,
\eeq
we get (assuming $m\leq n$)
\beqa\label{da}
& &\int_{-1}^1 \left(\calD_q^\kappa P_m^{(\a)}(x;\kappa)\right) P_n^{(\a)}(x;\kappa) \w^{(\a)}(x;\kappa) d_q x\nonumber\\
&+&\int_{-1}^1 P_m^{(\a)}(x;q\kappa) \left(\calD_q^\kappa P_n^{(\a)}(x;\kappa)\right) \w^{(\a)}(x;\kappa) d_q x\nonumber\\
&+&\int_{-1}^1 P_m^{(\a)}(x;q\kappa) P_n^{(\a)}(x;q\kappa) \left(\calD_q^\kappa \w^{(\a)}(x;\kappa)\right) d_q x=0.
\eeqa

Substituting the Fourier expansions, Eqs.\eqref{db} and \eqref{dc}, into the first two terms of \eqref{da}, and 
using the Pearson relation (in the $\kappa$ variable) for the $q$-Hermite weight, we get 
\beqa
 & &\d_{mn} \Xi_{nn}^{(\a)}+(1-\l \Xi_{mn}^{(\a)})\Xi_{mn}^{(\a)} 
-\l \sum_{j=0}^{m-1} \Xi_{jm}^{(\a)}\Xi_{jn}^{(\a)}\nonumber\\
         &+&\d_{mn} \left(\frac{q^2\kappa}{q-1}\right)\left\{\left(\bar A_{n+1}^{(\a)}\right)^2+\left(\bar A_n^{(\a)}\right)^2\right\}
         +\d_{m,n-2}\left(\frac{q^2 \kappa}{q-1}\right) \bar A_{n-1}^{(\a)}\bar A_n^{(\a)}=0.
\eeqa

In order to illustrate the content of this equation, we consider the following specializations:\

\begin{itemize}
\item[Case $1$:] $m<n-2 \Leftrightarrow m+3\leq n$\\
In this case, the master equation reduces to 
\beqs
(1-\l \Xi_{mm}^{(\a)}) \Xi_{mn}^{(\a)}-\l \sum_{j=0}^{m-1} \Xi_{j m}^{(\a)} \Xi_{j n}^{(\a)}=0.
\eeqs
By the mathematical induction, we show that $\Xi_{mn}^{(\a)}=0$, if $3\leq n-m$. Consequently, 
the Fourier expansion of the $q$-derivative (w.r.t $\kappa$ variable) on the $q$-Hermite 
orthonormal polynomials only consist of two terms:
\beq\label{}
\calD_q^\kappa P_n^{(\a)}(x,\kappa)=P_n^{(\a)}(x,\kappa) \Xi_{nn}^{(\a)}(\kappa)+P_{n-2}^{(\a)}(x,\kappa) \Xi_{n-2,n}^{(\a)}(\kappa).
\eeq

\item[Case $2$:] $m=n-2 \Leftrightarrow m-1=n-3$\\
In this case, the master equation reduces to 
\beqs
(1-\l \Xi_{n-2,n-2}^{(\a)}) \Xi_{n-2,n}^{(\a)}-\l \sum_{j=0}^{n-3} \Xi_{j,n-2}^{(\a)} \Xi_{j n}^{(\a)}+\frac{q^2 \kappa}{q-1}\bar A_n^{(\a)}\bar A_{n-1}^{(\a)}=0.
\eeqs
Since we have showed that $\Xi_{j n}^{(\a)}=0$ for $j\leq n-3$ in Case $1$, we can use the 
 equation above to express the off-diagonal Fourier coefficient in terms of the diagonal ones:
 \beq\label{5-2-1}
 \Xi_{n-1,n+1}^{(\a)}=\left(\frac{q^2 \kappa}{1-q}\right)\frac{\bar A_{n+1}^{(\a)} \bar A_n^{(\a)}}{1-\l \Xi_{n-1,n-1}^{(\a)}}=\left(\frac{q^2 \kappa}{1-q}\right)\frac{ A_{n+1}^{(\a)} A_n^{(\a)}}{1-\l \Xi_{n+1,n+1}^{(\a)}}. 
 \eeq
 For the second equality of the equation above, we use Eq.\eqref{5-2-3} to replace $\bar A_n^{(\a)}$ in terms of $A_n^{(\a)}$.
 
 \item[Case $3$:] $m=n-1,m-1=n-2$\\
 Due to the parity preserving property associated with the $q$-Hermite ensemble,\\ $\Xi_{n-1,n}^{(\a)}=0$, we 
 have no constraint in this case.\\
 
 \item[Case $4$:] $m=n$\\
 In this case, we have 
 \beq\label{}
 \Xi_{nn}^{(\a)}+(1-\l \Xi_{nn}^{(\a)})\Xi_{nn}^{(\a)}- \l(\Xi_{n-2,n}^{(\a)})^2=\left(\frac{q^2 \kappa}{1-q}\right)\left[\left(\bar A_{n+1}^{(\a)}\right)^2+\left(\bar A_n^{(\a)}\right)^2\right].
 \eeq  
 After suitable rearrangement, we get
 \beq\label{5-2-2}
 (1-\l \Xi_{nn}^{(\a)})^2+\frac{(q\kappa)^4\left(\bar A_n^{(\a)}\right)^2\left(\bar A_{n-1}^{(\a)}\right)^2}{ (1-\l \Xi_{n-2,n-2}^{(\a)})^2}
 =1-(q\kappa)^2\left\{\left(\bar A_{n+1}^{(\a)}\right)^2+\left(\bar A_n^{(\a)}\right)^2\right\}.
 \eeq
\end{itemize}
Similar to the case of the $q$-Laguerre ensemble Eq.\eqref{if}, we can solve $(1-\l \Xi_{nn}^{(\a)})^2$ as a 
continued fraction in terms of the recursive coefficient $\bar A_n^{(\a)}$.

On the other hand, by replacing $\bar A_n^{(\a)}$ into $A_n^{(\a)}$, using Eq.\eqref{5-2-3}, we can 
derive a quadratic equation for $(1-\l \Xi_{nn}^{(\a)})^2$,
 \beq\label{5-2-4}
 (1-\l \Xi_{nn}^{(\a)})^2+\frac{(q\kappa)^4\left(A_n^{(\a)}\right)^2\left( A_{n-1}^{(\a)}\right)^2}{ (1-\l \Xi_{nn}^{(\a)})^2}
 =1-(q\kappa)^2\left\{\left(\bar A_{n+1}^{(\a)}\right)^2+\left(\bar A_n^{(\a)}\right)^2\right\}.
 \eeq
Hence, $\Xi_{nn}^{(\a)}$ can be solved in terms of $A_n^{(\a)}$ and $\bar A_n^{(\a)}$, and 
substituting the solution of $\Xi_{nn}^{(\a)}$ for Eq.\eqref{5-2-4} back to Eq.\eqref{toda:3}, we get closed $q$-difference equations for the 
recursive coefficients of the generalized $q$-Hermite ensemble.

\subsection{On the compatibility of the quadratic relation and $q$-Toda equations}\
 \vspace{0.5cm}

In this section, we check the compatibility between the quadratic relation 
Eqs. \eqref{bc}, \eqref{bd},\eqref{be}, \eqref{bf} and the $q$-Toda equation Eqs. \eqref{toda:1}, \eqref{toda:2}, \eqref{toda:3}. To see this, we first example the 
Fourier coefficients of the $q$-derivative of the orthonormal $q$-Laguerre/Hermite 
polynomials (w.r.t $\kappa$). 

\bthm
The Fourier coefficients of the $q$-derivative (w.r.t $\kappa$) of the orthonormal 
$q$-Laguerre/Hermite polynomials (Eqs. \eqref{ia}, \eqref{db}) are related by the quadratic 
relations
\beqa
\Xi_{2n,2n}^{(\a)}(\kappa,q)&=&(1+q)\kappa\xi_{nn}^{(\a)}(\kappa^2,q^2),\nonumber\\
\Xi_{2n+1,2n+1}^{(\a)}(\kappa,q)&=&(1+q)\kappa\xi_{nn}^{(\a+1)}(\kappa^2,q^2).\nonumber
\eeqa
\ethm
\bpf
We relate the $q$-derivative (w.r.t $\kappa$) of the orthonormal $q$-Hermite 
polynomials Eq. \eqref{bf} to that of the $q$-Laguerre polynomials in two ways. 
First of all, for even polynomials
\begin{align}\label{5-3-1}
&\calD_q^\kappa P_{2n}^{(\a)}(x;\kappa,q)\nonumber\\
&=\frac{P_{2n}^{(\a)}(x;\kappa,q)-P_{2n}^{(\a)}(x;q\kappa,q)}{(1-q)\kappa}\nonumber\\
&=\frac{p_{n}^{(\a)}(x^2;\kappa^2,q^2)-p_{n}^{(\a)}(x^2;q^2\kappa^2,q^2)}{(1-q^2)\kappa^2}\frac{(1+q)^{\frac{3}{2}}\kappa}{\sqrt{2}}\nonumber\\
&=\frac{(1+q)^{\frac{3}{2}}\kappa}{\sqrt{2}}\left[\calD_{q^2}^{\kappa^2}p_{n}^{(\a)}(x^2;\kappa^2,q^2)\right]\nonumber\\
&=\frac{(1+q)^{\frac{3}{2}}\kappa}{\sqrt{2}}\left[p_{n}^{(\a)}(x^2;\kappa^2,q^2)\xi_{nn}^{(\a)}(\kappa^2,q^2)+p_{n-1}^{(\a)}(x^2;\kappa^2,q^2)\xi_{n-1,n}^{(\a)}(\kappa^2,q^2)\right].
\end{align}
On the other hand, if we write the Fourier expansion of the $q$-derivative (w.r.t $\kappa$) of the $q$-Hermite 
polynomials
\begin{align}\label{5-3-2}
&\calD_q^\kappa P_{2n}^{(\a)}(x;\kappa,q)\nonumber\\
&= P_{2n}^{(\a)}(x;\kappa,q) \Xi_{2n,2n}^{(\a)}(\kappa,q)+P_{2n-2}^{(\a)}(x;\kappa,q)\Xi_{2n-2,2n}^{(\a)}(\kappa,q)\nonumber\\
&=\sqrt{\frac{1+q}{2}}p_{n}^{(\a)}(x^2;\kappa^2,q^2)\Xi_{2n,2n}^{(\a)}(\kappa,q)+\sqrt{\frac{1+q}{2}}p_{n-1}^{(\a)}(x^2;\kappa^2,q^2)\Xi_{2n-2,2n}^{(\a)}(\kappa,q).
\end{align}
By comparing the two results Eqs. \eqref{5-3-1}, \eqref{5-3-2}, we obtain
\beqa
\Xi_{2n,2n}^{(\a)}(\kappa,q)&=&(1+q)\kappa \xi_{nn}^{(\a)}(\kappa^2,q^2),\label{5-3-5}\\
\Xi_{2n-2,2n}^{(\a)}(\kappa,q)&=&(1+q)\kappa \xi_{n-1,n}^{(\a)}(\kappa^2,q^2).\label{5-3-6}
\eeqa
Next, we compare the odd $q$-Hermite polynomials.
\begin{align}\label{5-3-3}
&\calD_q^\kappa P_{2n+1}^{(\a)}(x;\kappa,q)\nonumber\\
&=\frac{(1+q)^{\frac{3}{2}}\kappa}{\sqrt{2}} x \calD_{q^2}^{\kappa^2}p_{n}^{(\a+1)}(x^2;\kappa^2,q^2) \nonumber\\
&=\frac{(1+q)^{\frac{3}{2}}\kappa}{\sqrt{2}} x \left[p_{n}^{(\a+1)}(x^2;\kappa^2,q^2)\xi_{nn}^{(\a+1)}(\kappa^2,q^2)+p_{n-1}^{(\a+1)}(x^2;\kappa^2,q^2)\xi_{n-1,n}^{(\a+1)}(\kappa^2,q^2)\right],
\end{align}
and
\begin{align}\label{5-3-4}
&\calD_q^\kappa P_{2n+1}^{(\a)}(x;\kappa,q)\nonumber\\
&= P_{2n+1}^{(\a)}(x;\kappa,q) \Xi_{2n+1,2n+1}^{(\a)}(\kappa,q)+P_{2n-1}^{(\a)}(x;\kappa,q)\Xi_{2n-1,2n+1}^{(\a)}(\kappa,q)\nonumber\\
&=\sqrt{\frac{1+q}{2}}x p_{n}^{(\a+1)}(x^2;\kappa^2,q^2)\Xi_{2n+1,2n+1}^{(\a)}(\kappa,q)+\sqrt{\frac{1+q}{2}}x p_{n-1}^{(\a+1)}(x^2;\kappa^2,q^2)\Xi_{2n-1,2n+1}^{(\a)}(\kappa,q).
\end{align}
By comparing the two results, Eqs. \eqref{5-3-3}, \eqref{5-3-4}, we obtain
\beqa
\Xi_{2n+1,2n+1}^{(\a)}(\kappa,q)&=&(1+q)\kappa \xi_{nn}^{(\a+1)}(\kappa^2,q^2),\label{5-3-7}\\
\Xi_{2n-1,2n+1}^{(\a)}(\kappa,q)&=&(1+q)\kappa \xi_{n-1,n}^{(\a+1)}(\kappa^2,q^2).\label{5-3-8}
\eeqa
\epf

In fact, the quadratic relation among the Fourier coefficients of the $q$-Laguerre/Hermite 
polynomials are equivalent to the quadratic relation among the recursive coefficients of 
the $q$-Laguerre/Hermite polynomials. To see this, we rewrite the Eq. \eqref{5-2-1} (set $n=m-1$) as
\beq\label{}
A_m^{(\a)}(\kappa,q)A_{m-1}^{(\a)}(\kappa,q)=\frac{1-q}{q^2\kappa}\Xi_{m-2,m}^{(\a)}(\kappa,q)\left[1-(1-q)\kappa \Xi_{mm}^{(\a)}(\kappa,q)\right].
\eeq
For $m=2n$, after substituting Eqs. \eqref{5-3-5}, \eqref{5-3-6} and using Eq. \eqref{5-1-1}, we get
\beq\label{}
A_{2n}^{(\a)}(\kappa,q)A_{2n-1}^{(\a)}(\kappa,q)=\frac{1-q^2}{q^2}\xi_{n-1,n}^{(\a)}(\kappa^2,q^2)\left[1-(1-q^2)\kappa^2 \xi_{nn}^{(\a)}(\kappa^2,q^2)\right]=a_n^{(\a)}(\kappa^2,q^2).
\eeq 
For $m=2n+1$, after substituting Eqs. \eqref{5-3-7} , \eqref{5-3-8} and using Eq. \eqref{5-1-1}, we get
\beq\label{}
A_{2n+1}^{(\a)}(\kappa,q)A_{2n}^{(\a)}(\kappa,q)=\frac{1-q^2}{q^2}\xi_{n-1,n}^{(\a+1)}(\kappa^2,q^2)\left[1-(1-q^2)\kappa^2 \xi_{nn}^{(\a+1)}(\kappa^2,q^2)\right]=a_n^{(\a+1)}(\kappa^2,q^2).
\eeq 

Similarly, rewriting Eq. \eqref{5-2-2},
\begin{align}
&(q\kappa)^2\left[\left(A_{m+1}^{(\a)}(q\kappa,q)\right)^2+\left(A_{m}^{(\a)}(q\kappa,q)\right)^2\right]\nonumber\\
&=1-[1-(1-q)\kappa \Xi_{mm}(\kappa,q)]^2-\frac{(q\kappa)^4\left[A_{m}^{(\a)}(\kappa,q)\right]^2\left[A_{m-1}^{(\a)}(\kappa,q)\right]^2}{[1-(1-q)\kappa \Xi_{mm}(\kappa,q)]^2},
\end{align}
then by substituting Eqs. \eqref{bc}, \eqref{be}, \eqref{if}, \eqref{5-3-5}, \eqref{5-3-6}, for either $m=2n$ or $m=2n+1$, we 
reproduce Eqs. \eqref{bd}, \eqref{bf}.


\section{Summary and Conclusion}

In this paper, we study fundamental properties of the generalized
$q$-Laguerre/Hermite ensembles with a deformation parameter $\kappa$. 

For the special value $\kappa=1$, we give explicit solutions of their recursive 
coefficients for the orthonormal polynomial systems. In addition, to show that 
these ensembles provide ''good'' basis for a realization of the Heisenberg 
algebra, we also calculate the matrix elements (Fourier coefficients) of 
the dilation operator with respect ot orthonormal polynomial basis explicitly.

In view of the intimate connections between matrix models and orthogonal 
polynomial systems \cite{DiFrancesco:1993cyw}, we compute the Hankel determinants as products 
of the recursive coefficients explicitly. The physical consequences for 
identifying the Hankel determinants as partition functions of these matrix 
models with non-polynomial type potential may worth further explorations.

Finally, we examine the deformation of the $q$-Laguerre/Hermite systems 
in the form of $q$-difference equations dictating the $\kappa$ dependence 
of the recurrence coefficients and Fourier coefficients. This provides a $q$-generalization 
of the Toda equations in the Lax pair formulation \cite{F74}.

\section*{Acknowledgement}

This research project was initiated in a summer visit (by C. T.) to the Institute of Mathematics 
at Academia Sinica in 2015. Both authors would like to thank Derchyi Wu and 
Chueh-Hsin Chang, and Mourad E. H. Ismail for instructive discussions. C.T. 
would like to thank Derchyi Wu for invitation and hospitality. The research work 
of C.T. is partially supported by the grant from Academia Sinica for summer visit,
and in parts supported by the MOST research grant 104-2112-M-029-001. The 
research work of H.F. is partially supported by the MOST research 
grants 104-2115-M-001-001-MY2 and .

\newpage

\appendix
\section{Some Basic Definitions and Relations for 
              $q$-Analysis ($0<q<1$)}\label{app:a}



In this section, we collect some basic definitions and formulas which 
are relevant to our discussions.

The $q$-integral for a function $f(x)$ over the region $x\in [0,a]$ is defined as
\beq\label{}
F(a):=\int_0^a f(x) d_q x := a(1-q)\sum_{k=0}^\infty f(a q^k)q^k.
\eeq
This is compatible with the definition of the $q$-derivative
\beq\label{app:1}
\calD_q f(x) := \frac{f(q x)-f(x)}{qx-x}=\frac{f(x)-f(q x)}{x(1-q)}
\eeq
in the following senses:

\ben
\item Fundamental theorem of the $q$-calculus
\beqa
\calD_q F(a)&=&f(a),\\
\int_0^a [\calD_q f(x)] d_q x&=&f(a)-f(0).
\eeqa

\item The linear change of variables can be implemented in $q$-integral:
\beq\label{}
\int_0^a f(c x) d_q x=\frac{1}{c}\int_0^{ca} f(y) d_q y.
\eeq
\een

There are some subtleties associated with the $q$-derivative, in particular, the 
$q$-Lebinitz rule is given as
\beqa\label{app:2}
\calD_q[f(x)g(x)] & = & \frac{f(qx)g(qx)-f(x)g(x)}{(q-1)x} \nonumber\\
                           & = & f(qx)[\calD_q g(x)]+[\calD_q f(x)]g(x) \nonumber\\
                           & = & [\calD_q f(x)]g(qx)+f(x)[\calD_q g(x)].
\eeqa

\newpage
\section{On the $q$-deformation of the Heisenberg Algebra}

In the classical case, the Heisenberg algebra is generated by two operators:
$\calD:=\frac{d}{dx}$ (translation) and $x$ (position). They satisfy the commutation 
relation,
\beq
\left[\calD,x\right]=\left[\frac{d}{dx},x\right]=1.
\eeq

We can also include the dilation operator $\calS:=x \dfrac{d}{dx}$, such that
\beq
\left[\calS,x\right]=\left[x\frac{d}{dx},x\right]=x,
\eeq
and
\beq
\left[\calS,\calD\right]=\left[x\frac{d}{dx},\frac{d}{dx}\right]=-\frac{d}{dx}.
\eeq
In this way, we can view the position ($x$) and translation ($\frac{d}{dx}$) operators 
as raising and lowering operators with respect to the eigenfunctions of dilation. 

To study the $q$-deformation of the Heisenberg algebra, which is useful in our computations 
of the recursive coefficients for the $q$-Laguerre and $q$-Hermite ensembles, we
 compute the action of three commutators on a general function $f(x)$: ($\calS_q := x \calD_q$)
 
\bthm
The $q$-difference realization of the $q$-Heisenberg algebra is given by
\beqa
\left[\calD_q,x\right]f(x)&=&f(q x)\\
\left[\calS_q, x\right]f(x)&=&x f(q x)\\
\left[\calS_q,\calD_q\right]f(x)&=&-\calD_q f(q x).
\eeqa
\ethm
\begin{proof}

\beqas
[\calD_q,x] f(x) &=&\calD_q\left(x f(x)\right)-x\left(\calD_q f(x)\right)\\
           &=&\frac{qx f(qx)-x f(x)}{(q-1)x}-\frac{f(qx)-f(x)}{q-1}\\
           &=&f(qx).
\eeqas
\beqas
[\calS_q,x] f(x)&=&\calS_q\left(x f(x)\right)-x\left(\calS_q f(x)\right)\\
           &=&\frac{qx f(qx)-x f(x)}{q-1}-x \frac{f(qx)-f(x)}{q-1}\\
           &=&x f(qx).
\eeqas
\beqa
[\calS_q,\calD_q] f(x)&=&\calS_q\left(\calD_q f(x)\right)-\calD_q\left(\calS_q f(x)\right)\nonumber\\
           &=&\calS_q\left(\frac{f(qx)- f(x)}{(q-1)x}\right)-\calD_q\left( \frac{f(qx)-f(x)}{q-1}\right)\nonumber\\
           &=&-\frac{f(q^2x)}{q(q-1)x}+\frac{f(qx)}{q(q-1)x}\nonumber\\
           &=&-\calD_q f(qx).
\eeqa
\end{proof}

\bthm
The eigenfunctions of the $q$-dilation operator are monomial $f_n(x)=x^n$ with eigenvalue 
$\dfrac{1-q^n}{1-q}$.
\ethm
\bpf
\beqs
\calS_q f_n(x)=x\left[\calD_q f_n(x)\right]=\frac{x^n-q^n x^n}{1-q}=\left(\frac{1-q^n}{1-q}\right)x^n.
\eeqs
\epf

\bthm
For any $q$-weight function, the diagonal matrix elements (or the Fourier coefficients) of 
the $q$-dilation operator in terms of orthonormal polynomial basis, defined by
\beqs
\calS_q p_n(x)=\sum_{j=0}^n p_j(x)c_{j n},
\eeqs
have an universal expression:
\beq
c_{nn}=\frac{1-q^n}{1-q}.
\eeq
\ethm
\bpf
By using $[\calD_q, x]p_n(x)=p_n(q x)$, we express
\beqs
LHS=\frac{xp_n(x)-(qx)p_n(qx)}{(1-q)x}-(c_{nn}p_n+c_{n-1,n}p_{n-1}+\cdots)
\eeqs
and
\beqs
RHS=\left[\g_n(qx)^n+\d_n(qx)^{n-1}+\cdots\right].
\eeqs
Then we compare the coefficient of $x^{n}$ to get
\beqs
\frac{\g_n-q^{n+1}\g_n}{1-q}-\g_n c_{nn}=q^n\g_n,
\eeqs
which implies
\beqs
c_{nn}=\frac{1-q^{n+1}}{1-q}-q^n=\frac{1-q^{n}}{1-q}.
\eeqs
\epf

\newpage
\section{A Short Note for the $q$-exponential function}
The $q$-deformed exponential function is defined either by an infinite 
product or as a power series in $x$,

\begin{align}
e_q(x)&:=\frac{1}{\prod_{k=0}^\infty(1-q^k x)}\\
          &=1+\sum_{n=1}^\infty \frac{x^n}{\prod_{k=1}^n (1-q^k)}\\
          &=1+\sum_{n=1}^\infty \frac{x^n}{(1-q)(1-q^2)\cdots(1-q^n)}. \label{app:c1}
\end{align}

\bthm
The $q$-deformed exponential function satisfies the following $q$-difference equation:
\beq
\calD_q^x e_q(x)=\left(\frac{1}{1-q}\right)e_q(x).
\eeq
\ethm
\bpf
\begin{align*}
&\calD_q^x e_q(x)\\
&=\frac{e_q(x)-e_q(qx)}{(1-q)x}\\
&=\frac{1}{(1-q)x}\left[\frac{x}{1-q}+\sum_{n=2}^\infty \frac{x^n}{(1-q)(1-q^2)\cdots (1-q^n)} -\frac{qx}{1-q}-\sum_{n=2}^\infty\frac{q^nx^n}{(1-q)(1-q^2)\cdots(1-q^n)}\right]\\
&=\left(\frac{1}{(1-q)}\right)\left[1+\sum_{n=2}^\infty \frac{x^n-1}{(1-q)(1-q^2)\cdots (1-q^{n-1})}\right]\\
&=\left(\frac{1}{(1-q)}\right)e_q(x).
\end{align*}
\epf
From this result, it is easy to see that, using the power series form, Eq. \eqref{app:c1}, as $\e := 1-q \rightarrow 0$, 
we can connect this $q$-exponential function to 
the classical exponential and Gaussian functions.
\beqs
\lim_{\e \rightarrow 0} e_{1-\e}(\e x)= \lim_{\e \rightarrow 0} \left(1+\sum_{n=1}^\infty \frac{\e^n x^n}{\e \cdot 2\e \cdot \cdots \cdot n\e}\right)=e^x.
\eeqs

\beq\label{}
\lim_{q\rightarrow 1} e_{q^2} ((1-q^2)x^2)= \lim_{\e \rightarrow 0} \left(1+\sum_{n=1}^\infty \frac{(2\e)^n x^{2n}}{2\e \cdot 4\e \cdot \cdots \cdot 2n\e}\right)=e^{x^2}.
\eeq

\bthm
The general eigenfunctions of the $q$-derivative operator are $e_q((1-q)\kappa x)$ (up to normalization) 
with eigenvalue $\kappa$.
\ethm
\bpf
We can solve the eigenvalue problem
\beq\label{app:c1}
\calD_q f_\kappa(x)=\kappa f_\kappa(x)
\eeq
by using power series expansion, let
\beqa
f_\kappa(x)&=&c_0+\sum_{n=1}^\infty c_n x^n, \label{app:c2}\\
\Rightarrow f_\kappa(qx)&=&c_0+\sum_{n=1}^\infty c_n q^n x^n, \nonumber\\
\calD_q f_\kappa(x)&=& \frac{f_\kappa(x)-f_\kappa(qx)}{(1-q)x} \nonumber\\
&=&c_1+\sum_{n=1}^\infty c_{n+1}\left(\frac{1-q^{n+1}}{1-q}\right) x^n. \label{app:c3}
\eeqa
\epf

By comparing Eq. \eqref{app:c3} with Eq. \eqref{app:c1}, we obtain a recursive relation 
among the coefficients $c_n$
\beqs
c_n=\frac{(1-q)\kappa}{1-q^n}c_{n-1}.
\eeqs
Telescoping leads to general solution 
\beqs
c_n=\frac{(1-q)^n\kappa^n}{\prod_{l=1}^n (1-q^l)}c_0,
\eeqs
and $f_\kappa(x)=c_0 e_q((1-q)\kappa x)$.\

Equipped with these preliminaries, we can compute the zeroth moment of the 
$q$-Laguerre ensemble (with $\a=0$).

\bthm\label{thm:c1}
The zeroth moment of the $q$-Laguerre ensemble is given by
\beq\label{app:c4}
\mu_{0,L}^{(0)}=\left(\frac{1-q}{q\kappa}\right)\left[1-(\kappa;q)_\infty\right].
\eeq
\ethm
\bpf
In this paper, the $q$-Laguerre weight function is defined as the inverse 
of the $q$-Pochhammer symbol,
\beq
e_q(\kappa x) v^{(0)}(x;\frac{\kappa}{q},q)=1.
\eeq
By taking $q$-derivative on both sides of the equation above, we get
\beq
\left[\calD_q^x e_q(\kappa x)\right] v^{(0)}(x;\frac{\kappa}{q},q)+e_q(q\kappa x)\left[\calD_q^x v^{(0)}(x;\frac{\kappa}{q},q)\right]=0.
\eeq
Thus, the $\a=0$ $q$-Laguerre weight can be written as a total $q$-derivative,
\beqa
v^{(0)}(x;\kappa,q)&=&\frac{1}{e_q(q\kappa x)}=(q\kappa x;q)_\infty \nonumber\\
&=&\frac{-1}{\left[\calD_q^x e_q(\kappa x)\right] v^{(0)}(x;\frac{\kappa}{q},q)}\calD_q^x v^{(0)}(x;\frac{\kappa}{q},q) \nonumber\\
&=&\frac{-(1-q)}{\kappa}\calD_q^x (\kappa x;q)_\infty. \label{app:c4}
\eeqa
Using the fundamental theorem of the $q$-calculus, the zeroth moment can be calculated as
\beqa
\mu_{0,L}^{(0)}&=&\int_0^1 v^{(0)}(x;\kappa,q) d_q x\nonumber\\
&=&-\left(\frac{1-q}{q\kappa}\right)\int_0^1 \calD_q^x (\kappa x;q)_\infty d_q x \nonumber\\
&=&\left(\frac{1-q}{q\kappa}\right)\left[1-(\kappa;q)_\infty\right]. \label{app:c5}
\eeqa
\epf

Similarly, we can compute the zeroth moment of the $q$-Hermite ensemble.
\bthm\label{thm:c4}
The zeroth moment of the $q$-Hermite ensemble is given by the incomplete 
$q$-Gamma integral
\beq\label{app:c6}
\mu_{0,H}^{(-\frac{1}{2})}=\frac{2}{1+q}\int_0^1 y^{-\frac{1}{2}}(q^2\kappa^2 y;q^2)_\infty d_{q^2}y.
\eeq
\ethm
\bpf
Using the formula of change of variables for $q$-integral ($y=x^2$)
\beq\label{app:c7}
\int_0^1 f(x^2)d_q x=\frac{1}{1+q} \int_0^1 \frac{f(y)}{\sqrt{y}} d_{q^2}y,
\eeq
we can compute the zeroth moment of the $q$-Hermite ensemble (with $\a=-\frac{1}{2}$)
\beqa
\mu_{0,H}^{(-\frac{1}{2})}(\kappa,q)&=& 2 \int_0^1 (q^2\kappa^2x^2;q^2)_\infty d_q x \nonumber\\
&=&\frac{2}{1+q} \int_0^1 y^{-\frac{1}{2}} (q^2\kappa^2 y;q^2)_\infty d_{q^2}y. \label{app:c8}
\eeqa
\epf
By comparing with the integral representation of the $q$-Gamma function \cite{SK05},
\beqa
\Gamma_q(t)&:=&(1-q)^{-t}\int_0^1 x^{t-1}(qx;q)_\infty d_q x,\\
\Gamma_{q^2}(\frac{1}{2})&=&\frac{1}{(1-q^2)^{\frac{1}{2}}} \int_0^1 \frac{1}{\sqrt{x}} (q^2x;q^2)_\infty d_{q^2} x\nonumber\\
&=&\frac{\kappa}{(1-q^2)^{\frac{1}{2}}} \int_0^{\kappa^2} \frac{1}{\sqrt{y}} (q^2 \kappa^2 y; q^2)_\infty d_{q^2} y,
\eeqa
we find that 
\beq
\mu_{0,H}^{(-\frac{1}{2})}(1,q)=2 \sqrt{\frac{1-q}{1+q}}\Gamma_{q^2}(\frac{1}{2}).
\eeq
In addition, there exists a quadratic relation between the zeroth weights of $q$-Laguerre/Hermite ensembles
\beq
\mu_{0,H}^{(-\frac{1}{2})}(\kappa,q)=\frac{2}{1+q}\mu_{0,L}^{(-\frac{1}{2})}(\kappa^2,q^2).
\eeq

\newpage
\section{Matrix elements of the translation operator}

In this appendix, we calculate the matrix elements of the translation operators 
in terms of recursive coefficients.

\subsection{Matrix elements of the translation operator in the basis of Laguerre polynomials}\

The matrix elements of the translation operator in the basis of Laguerre polynomials are 
defined as
\beq
\calD p_n(x) =\sum_{j=0}^{n-1} p_j(x) t_{jn}.
\eeq
By multiplying both sides of the equation by position operator, we get 
\beq
\calS p_n(x)= x \calD p_n(x) = \sum_{j=0}^{n-1}\left[xp_j(x)\right] t_{jn}.
\eeq
Using the result in Sec. \ref{sec:3-1}, Eq. \eqref{3-1-2}, we get
\beq
c_{nn}p_n(x)+c_{n-1,n}p_{n-1}(x)= \sum_{j=0}^{n-1}\left[a_{j+1}p_{j+1}(x)+b_j p_j(x)+a_jp_{j-1}(x)\right] t_{jn}.
\eeq
Comparing the coefficients on both sides, we get a set of coupled equations for  $t_{mn}$:
\beqas
c_{nn}&=&a_nt_{n-1,n},\\
c_{n-1,n}&=&a_{n-1}t_{n-2,n}+b_{n-1}t_{n-1,n},\\
0=c_{n-k,n}&=&a_{n-k}t_{n-k-1,n}+b_{n-k}t_{n-k,n}+a_{n-k+1}t_{n-k+1,n}. \quad (k \geq 2)
\eeqas
The solution can be solved by iteration and are given by
\beqas
t_{n-1,n}&=&\frac{c_{nn}}{a_n},\\
t_{n-2,n}&=&\frac{1}{a_{n-1}}\left[c_{n-1,n}-\frac{b_{n-1}c_{nn}}{a_n}\right],\\
t_{n-k-1,n}&=&-\frac{b_{n-k}}{a_{n-k}}t_{n-k,n}-\frac{a_{n-k+1}}{a_{n-k}}t_{n-k+1,n}. \quad (k \geq 2)
\eeqas

\subsection{Matrix elements of the translation operator in the basis of Hermite polynomials}\

The matrix elements of the translation operator in the basis of Hermite polynomials are 
defined as
\beq
\calD P_n(x) =\sum_{j=0}^{n-1} P_j(x) T_{jn}. \quad n-j: \mbox{ odd}
\eeq
By multiplying both sides of the equation by position operator, we get 
\beq
\calS P_n(x)= x \calD P_n(x) = \sum_{j=0}^{n-1}\left[xP_j(x)\right] T_{jn}.
\eeq
Using the result in Sec. \ref{sec:3-2}, Eq.\eqref{3-2-3}, we get
\beq
C_{nn}P_n(x)+C_{n-2,n}P_{n-2}(x)= \sum_{j=0}^{n-1}\left[A_{j+1}P_{j+1}(x)+A_jP_{j-1}(x)\right] T_{jn}.
\eeq
Comparing the coefficients on both sides, we get a set of coupled equations for  $T_{mn}$:
\beqas
C_{nn}&=&A_nT_{n-1,n},\\
C_{n-2,n}&=&A_{n-2}T_{n-3,n}+A_{n-1}T_{n-1,n},\\
0=C_{n-k,n}&=&A_{n-k}T_{n-k-1,n}+A_{n-k+1}T_{n-k+1,n}. \quad (k \geq 4)
\eeqas
The solution can be solved by iteration and are given by
\beqas
T_{n-1,n}&=&\frac{C_{nn}}{A_n},\\
T_{n-(2k+1),n}&=&(-1)^k\frac{A_{n-2k+1}A_{n-2k-1}\cdots A_{n-1}C_{nn}}{A_{n-2k}A_{n-2k-2}\cdots A_{n-2}A_n}. \quad (k \geq 1)
\eeqas

\bibliographystyle{plain}

\end{document}